\documentstyle[twocolumn,prd,aps,epsf]{revtex}
\begin{document}
\def \slash#1{\not\! #1}
\def \r#1{(\ref{#1})}
\def \inp#1#2{#1\!\cdot\!#2}
\def \inperp#1#2{#1^\perp\!\!\cdot\!#2^\perp}
\def \cd{\makebox[0.08cm]{$\cdot$}}
\def \Mone  {\raisebox{-0.25cm}{\epsfxsize=1.6cm \epsffile{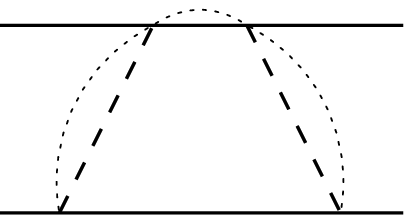}}}
\def \Mtwo  {\raisebox{-0.25cm}{\epsfxsize=1.6cm \epsffile{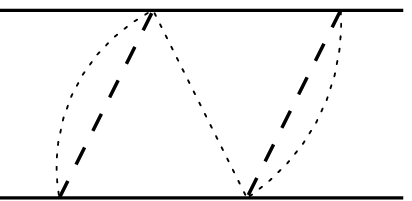}}}
\def \Mthree{\raisebox{-0.25cm}{\epsfxsize=1.6cm \epsffile{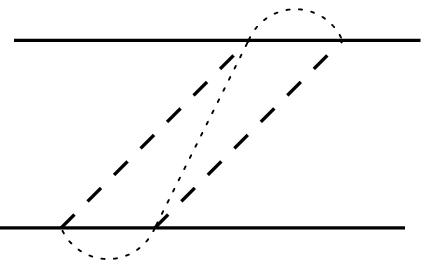}}}
\draft

\twocolumn[\hsize\textwidth\columnwidth\hsize\csname
@twocolumnfalse\endcsname

\title{Entanglement of Fock-space expansion and covariance in   
light-front Hamiltonian dynamics$^*$}
\author{N.C.J. Schoonderwoerd$^\dag$, B.L.G. Bakker$^\ddag$\\
{\em Department of Physics and Astronomy, Vrije Universiteit, Amsterdam,
The Netherlands, 1081 HV}
\\[.3cm] V.A. Karmanov$^\S$\\
{\em Lebedev Physical Institute, Leninsky Prospekt 53, 117924 Moscow, Russia}
\\[.3cm]}
\date{12 June 1998}
\maketitle

\begin{abstract}
We investigate in a model with scalar ``nucleons'' and mesons the
contributions of higher Fock states that are neglected in the ladder
approximation of the Lippmann-Schwinger equation. This leads to a
breaking of covariance, both in light-front and in instant-form
Hamiltonian dynamics. The lowest Fock sector neglected has two mesons
in the intermediate state and corresponds to the stretched box.  First
we show in a simplified example that the contributions of higher Fock
states are much smaller on the light-front than in instant-form dynamics.
Then we show for a scattering amplitude above threshold that the
stretched boxes are small, however, necessary to retain covariance. For
an off energy-shell amplitude covariance is not necessarily maintained
and this is confirmed by our calculations. Again, the stretched boxes
are found to be small. 
\\
\end{abstract}
]
\addtocounter{footnote}{1}
\footnotetext{VUTH 98-20, submitted to Phys.Rev.C}
\addtocounter{footnote}{1}
\footnotetext{nico@nat.vu.nl}
\addtocounter{footnote}{1}
\footnotetext{blgbkkr@nat.vu.nl}
\addtocounter{footnote}{1}
\footnotetext{karmanov@sci.lebedev.ru}

\section{Introduction}

In his famous article of 1949, Dirac \cite{Dir49} described a number of
ways how to set up a framework of Hamiltonian dynamics. Two of these
are most important. In instant-form (IF) Hamiltonian dynamics one
specifies the initial conditions on the equal-time plane $t=0$.  Of the
ten Poincar\'{e} generators six are kinematic, i.e., do not contain 
interactions, and are therefore conserved
quantities, and four are dynamical: the three boost operators and the
energy. In light-front (LF) Hamiltonian dynamics one uses the coordinates

\begin{equation}
A^\perp  =  (A^x, A^y), \quad A^\pm = \frac{A^0 \pm A^z}{\sqrt{2}},
\end{equation}
and quantizes on the plane $x^+=0$. Then only three operators are
dynamical. Two of them involve rotations around the $x$ or $y$-axis.
Therefore LF time-ordered amplitudes are not invariant under such
rotations.

The question of rotations in LF dynamics (LFD) was discussed before
with the aim of constructing the angular momentum operators, see e.g.
Fuda~\cite{Fud91}, and the review by Carbonell {\em et al.} \cite{CDKM98}.
While these authors emphasize the algebraic properties of the
generators of the Poincar\'{e} group, we stress in the present paper
the connection between expansions in Fock-space and covariance. It has
been remarked before by Brodsky {\em et al.} \cite{BJS85} that the
higher components in Fock space contribute to the difference between
the Bethe-Salpeter equation and the evolution equation in LFD. These
authors do not give numerical estimates of the corrections. The latter
has been done by Mangin-Brinet and Carbonell \cite{MC97}, and by Frederico
{\em et al.} \cite{SFC98}, who studied the same model and found the
effect of higher Fock states on the binding energy to be small.
In a calculation of positronium, Trittmann and Pauli \cite{TP97}  used
an effective theory, where the effects of all Fock states are included
in the interaction. They found rotational symmetry to be restored in the
solution.

In this paper we
consider first standard LF quantization and discuss the problem of
noncovariance, which includes violation of rotational invariance, in
the framework of LF time-ordered perturbation theory. We give numerical
results for the simplified case of two heavy scalars exchanging light
scalar particles. This choice is motivated by the popular
meson-exchange models in nuclear physics. We do not include the
internal spin degrees of freedom, as this is a complication that may
obscure the main point of our investigation: the connection between the
breaking of covariance and a truncation of the expansion in Fock
space.  In two interesting papers, Fuda \cite{Fud95f,Fud95i} reported
on detailed calculations of realistic one-meson exchange models in both
LF and IF dynamics. There the emphasis is on the comparison between the
two, when in both cases the ladder approximation is made. It is the
purpose of the present paper to show to what extent the ladder
approximation violates covariance.

\subsection{Suppression of higher Fock states}
A reason why LFD is preferred by many is that
higher Fock states are said to be more strongly suppressed in this form
of dynamics. As a disadvantage the lack of manifest rotational
invariance, and therefore covariance, is mentioned. We call a symmetry
manifest when it is connected to a kinematical operator. Then all
time-ordered diagrams exhibit this symmetry. Equal-time ordered
diagrams lack boost invariance whereas on the LF the
longitudinal boost $P^+$ is a kinematical operator. Therefore, if one
complains about lack of manifest covariance, one should include not
only rotational invariance but also other nonmanifest symmetries. 
One reason why people have rather stressed rotational invariance comes
easily to mind: in many cases it is easy to convince oneself by
inspection whether a matrix element is rotational invariant, viz when
the amplitude can be expressed in terms of scalar 
products of three-vectors. On the other hand, it is not more difficult
to test numerically for invariance under boost transformations than for
invariance under rotations. Indeed, the method used in the present
paper can easily be extended to check for boost invariance.

A way to test for covariance is to compare the LF time-ordered diagrams
to the covariant amplitude, since we know that the latter is invariant
under any of the Poincar\'{e} symmetry operations.  For on energy-shell 
amplitudes ($S$-matrix elements) there is an exact equality, as was proved by
Ligterink and Bakker \cite{LB95b} and which is confirmed in our
results. Off energy-shell there is a deviation, which, however, is
found in the present paper to be surprisingly small.

So why are we using these LF time-ordered diagram in the first place,
when there is an equivalent covariant method available? We do, because
we want to determine the properties of the bound state using the
Hamiltonian form of dynamics. In this method, covariance can never be
fully maintained. However, one may try to apply it in such a way that
breaking of covariance is minimal.  In many applications in nuclear
physics a one-meson exchange approximation is made for the interaction
and the scattering amplitude is computed by formally iterating this
interaction, leading to the Lippmann-Schwinger equation in the ladder
approximation. In this approximation one retains two- and
three-particle intermediate state and  neglects Fock states containing
four or more particles. These Fock sectors are needed to make the sum
of LF time-ordered diagrams equal to the covariant amplitude,
exhibiting the symmetries under all Poincar\'{e} transformations. If
these contributions are large, one can expect a significant breaking of
covariance, since the LF time-ordered diagrams are only invariant under
application of the kinematical symmetries.

For this reason we concentrate in this article on the determination of
the contributions of these higher Fock states. Our main concern shall be
the box diagram. Then we label the correction as ${\cal R}_{4^+}$. 
We shall calculate ${\cal R}_{4^+}$ explicitly for the box diagram with
scalar particles of different masses. The box diagram can be associated
with the two-meson exchange between two nucleons.  If spin would be
included several well-known complications would arise, the most
important one being the occurrence of instantaneous propagators
\cite{LB95b,KS70}. We do not want these complications to interfere with
the main point of our investigation:  the connection between Fock-space
truncations and lack of covariance.  Therefore spin is omitted.  We
have not included crossed box diagrams, because they are not relevant for
a discussion on covariance, since both the crossed and noncrossed box
diagram are covariant by themselves. 

\subsection{Setup}
First we explain the Lippmann-Schwinger formalism and the special
role of the box diagram. In Sec.~\ref{secbox} we describe how to
calculate both the covariant and the LF time-ordered amplitudes. After
this, we are ready for our numerical experiments.
In Sec.~\ref{secversus} the masses of the external particles are chosen
in such a way that on-shell singularities of the intermediate states
are avoided, and therefore it is easy to compare IF and LF Hamiltonian
dynamics. In that section it is shown that ${\cal R}_{4^+}$ is much
smaller in LFD than in IF dynamics (IFD), confirming the claim that in
LFD higher Fock states are more strongly suppressed. Moreover, it tells
us that covariance is more vulnerable in IFD than on the light-front!

After this exercise, we concentrate on the LF, and in
Sec.~\ref{secabovet} we calculate the LF time-ordered diagrams
for the more interesting case in which we have particles of fixed masses $m$
(called nucleons) and $\mu$ (called mesons). As the process we are concerned 
with, scattering, is above threshold, we have to deal with on-shell
singularities. We show that the breaking of covariance is again small.

Although in Sec.~\ref{secundert}, where we discuss off-shell amplitudes below
threshold, no on-shell singularities are encountered,
matters become more complicated because the notion of the c.m. frame
becomes ambiguous, since the total momentum $P^z$ is dynamical and
found to be unequal to the combined momentum of the two particles, $p^z
+ q^z$. However, we are still able to relate the breaking of covariance and
Fock-space truncation.

The lack covariance of the LF time-ordered amplitudes means that the
amplitude depends not only on the scalar products of the external
momenta, but on the angles between the quantization axis and the
external momenta as well.  Consequently, the amplitudes must have
singularities as a function of these angles in addition to the
familiar singularities as functions of the invariants.  The positions
of these singularities are found analytically in Sec.~\ref{seconshell},
in the framework of explicitly covariant LFD.  This gives a qualitative
understanding of the numerical results in Secs.~\ref{secversus}
and~\ref{secabovet}. In Sec.~\ref{secoffshell} explicitly covariant LFD
is applied to the off energy-shell results of Sec.~\ref{secundert}.

\section{The Lippmann-Schwinger formalism}
\label{seclippmann}
The Hamiltonian method aims at the determination of stationary states, i.e.
eigenstates of the Hamiltonian. Here we take the Yukawa-type model with
scalar coupling
\begin{equation}
 {\cal L}_{\rm int} = g \Phi^2 \phi .
 \label{eq.lag}
\end{equation}
Two types of particles are considered:  ``nucleons'' (N, $\Phi$) with
mass $m$ and ``mesons'' (m, $\phi$) with mass $\mu$.  The Hamiltonian
$H\equiv P^-$ consists of a part $H_0$ which describes free particles
and a part $V$ which describes the interaction:
\begin{equation}
H = H_0 + V.
\end{equation}
We shall denote the second term on the r.h.s. as the potential.
The problem of constructing the Hamiltonian from the underlying
Lagrangian has been recently reviewed by Brodsky {\em et al.} \cite{BPP97}.
Here we study two-nucleon states only. Moreover, we neglect self-energy 
diagrams.

We consider $H_0$, the kinematic part of the Hamiltonian in the two-nucleon 
(2N) sector. In the instant-form, quantization is carried out on planes of 
constant time ({\em equal-time planes}). 
Then we find for two particles of mass $m$ and momenta $p$
and $q$ respectively
\begin{equation}
H_0^{\rm IF} = \sqrt{\vec{p}^2 + m^2} + \sqrt{\vec{q}^2 + m^2},
\end{equation}
which leads to both negative and positive energy solutions. 
It is well-known \cite{BKT79}
that in this form the overall momenta and the relative momenta are difficult to
separate. 

In LF quantization the square root, and therefore the negative energy
solutions are absent. 
The interaction-free part of the two-body Hamiltonian is
\begin{equation}
H_0^{\rm LF} = \frac{{p^\perp}^2 + m^2}{2 p^+} +
      \frac{{q^\perp}^2 + m^2}{2 q^+} ,
\end{equation}
which demonstrates that positive energies occur for positive plus-momenta.
Moreover, one can easily separate the motion of a many-particle system as a 
whole from the internal motion of its constituents in the LF case \cite{BKT79}.

We shall focus on light-front quantization of our model in which the 
interaction of the nucleons is due to meson exchange.  We write the potential
in the form
\begin{equation}
\label{Vexp}
V = V_1 + V_2 + V_3 + \dots
\end{equation}
where the subscript denotes the number of mesons simultaneously exchanged. 
The potentials only contain irreducible diagrams to prevent
double counting. $V_1$ contains one-meson exchanges only:
\begin{equation}
\label{obep}
V_1 = 
   \raisebox{-0.45cm}{\epsfxsize=2.2cm \epsffile{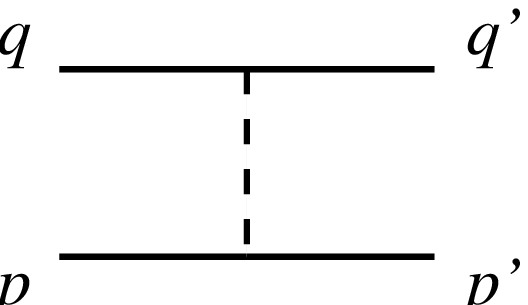}} 
 = \raisebox{-0.25cm}{\epsfxsize=1.7cm \epsffile{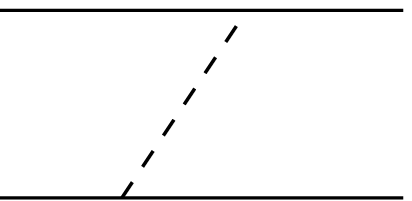}} +
   \raisebox{-0.25cm}{\epsfxsize=1.7cm \epsffile{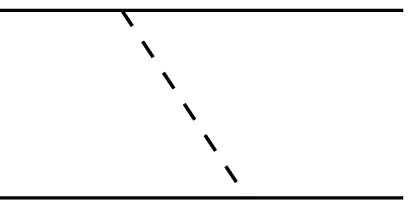}}
\end{equation}
The irreducible diagrams contributing to $V_1$ are depicted in
Eq.~\r{obep}. 
In these diagrams time goes from left to right. The nucleons are denoted
by solid lines, and the mesons by dashed lines.
Irreducible diagrams contributing to $V_2$ are those diagrams 
of order $g^4$ that cannot be
separated into two pieces by cutting two nucleon lines or two nucleon lines and
one meson line only. In terms of Fock-space sectors this means that $V_1$
contains two-nucleon and one-meson intermediate states, and $V_2$ contains only
two-nucleon two-meson intermediate states.

The potential $V_1$ is a covariant object in the case
the external lines are on shell.
The meaning of the equal sign in Eq.~\r{obep} is that the full
covariant amplitude can be written as a sum of two LF time-ordered diagrams.
Whereas the Feynman diagram contains the propagator $1/((q'-q)^2 - \mu^2)$, the
LF time-ordered diagrams contain the energy denominator $1/(P^- - H_0)$, 
$P^-$ being the parametric energy.  $H_0$ is the sum of the kinetic energies 
of the particles in the intermediate state: 
\begin{equation}
 H_0 = \sum_i \frac{{k^\perp_{i}}^2 + m^2_i}{2k^+_i} .
 \label{eq.ben1}
\end{equation}
The two diagrams contain $\theta$-functions of the plus-component of the 
momentum of the exchanged meson: one has the factor $\theta(p^+ - q^+)$, the
other $\theta(q^+ - p^+)$.

In a Feynman diagram the external lines are on mass shell and the initial
and final states have the same energy, which coincides with the parametric
energy. Then the minus component of the total four-momentum of a two-particle state
satisfies the relation
\begin{equation}
 P^- = p^- + q^- = p'^- + q'^- =
 \frac{p^2_\perp + m^2}{2 p^+} + \frac{q^2_\perp + m^2}{2 q^+} .
 \label{eq.ben2}
\end{equation}
As the minus component of the total momentum is the only dynamical momentum
operator, the other three components are conserved in any LF time-ordered
diagram. For instance, $P^+ = p^+ + q^+ = p'^+ + q'^+$. This conservation law
is very important in LF quantization. It leads immediately to the
{\em spectrum condition}: in any intermediate state all massive particles have
plus-momenta greater than zero and the sum of the plus-components of
the momenta of the particles in that state is equal to the total plus-momentum.

The expansion in Fock space does not coincide with an expansion in powers of 
the coupling constant. This can easily be seen when one considers an approach
closely resembling the Lippmann-Schwinger method.
The eigenstates $|\psi>$ of the Hamiltonian
\begin{equation}
 H |\psi> = P^- |\psi>,
\end{equation}
are also solutions of the Lippmann-Schwinger equation
\begin{equation}
 | \psi \rangle = | \phi \rangle + \frac{1}{P^- - H_0} V | \psi \rangle .
 \label{eq.ben3}
\end{equation}
The  formal solution of this equation is
\begin{equation}
 |\psi> = \sum_{i=0}^\infty \left( \frac{1}{P^- - H_0} V \right)^i |\phi>.
\end{equation}
 
An equation similar to Eq.~(\ref{eq.ben3}) exists for the scattering amplitude:
\begin{equation}
\raisebox{-0.25cm}{\epsfysize=0.65cm \epsffile{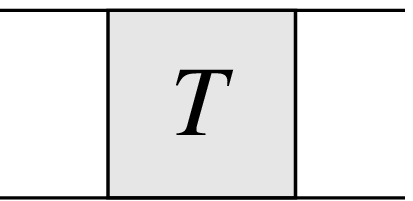}} =
\raisebox{-0.25cm}{\epsfysize=0.65cm \epsffile{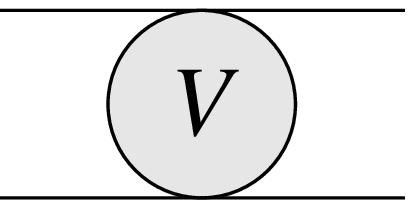}} +
\raisebox{-0.25cm}{\epsfysize=0.65cm \epsffile{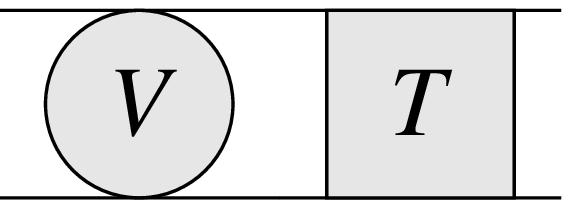}} 
\end{equation}
If one substitutes $V_1$ for $V$ in these equations one obtains the {\em ladder
approximation}. This approximation does not generate all diagrams, so one needs
to add corrections. At order $g^4$ this correction is $V_2$:

\begin{eqnarray}
V_2 &=&\raisebox{-0.25cm}{\epsfxsize=1.4cm \epsffile{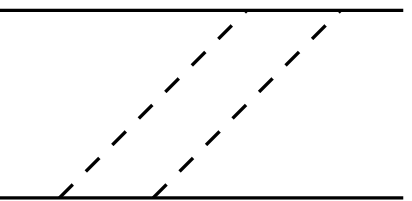}} +
       \raisebox{-0.25cm}{\epsfxsize=1.4cm \epsffile{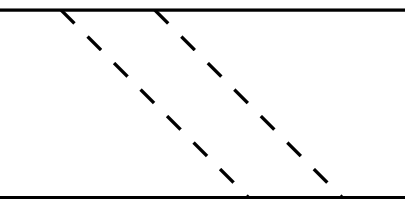}} 
\end{eqnarray}
If one takes into account all the contributions to $V$ from Eq.~\r{Vexp} then
the full scattering amplitude is
\begin{eqnarray}
  \raisebox{-0.25cm}{\epsfysize=0.65cm\epsffile{T.eps}} =
  \raisebox{-0.25cm}{\epsfxsize=1.4cm \epsffile{oneba.eps}} &+&
  \raisebox{-0.25cm}{\epsfxsize=1.4cm \epsffile{onebb.eps}} \nonumber\\
+ \raisebox{-0.25cm}{\epsfxsize=1.4cm \epsffile{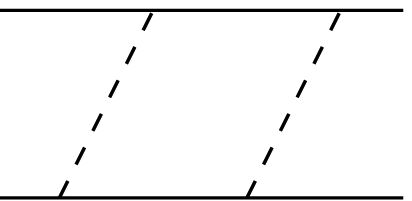}} +
  \raisebox{-0.25cm}{\epsfxsize=1.4cm \epsffile{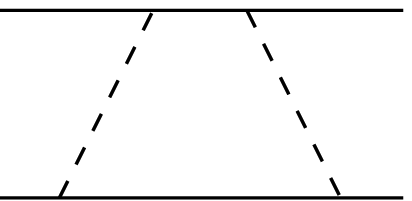}} &+&
  \raisebox{-0.25cm}{\epsfxsize=1.4cm \epsffile{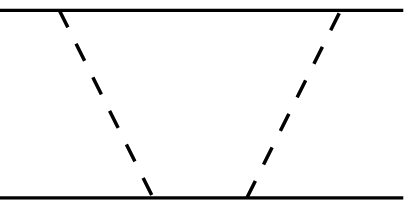}} +
  \raisebox{-0.25cm}{\epsfxsize=1.4cm \epsffile{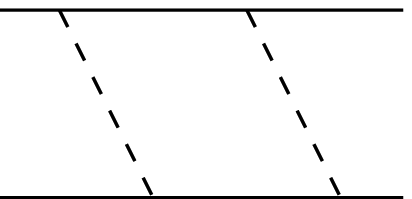}} \nonumber\\
+ \raisebox{-0.25cm}{\epsfxsize=1.4cm \epsffile{box0.eps}} + 
  \raisebox{-0.25cm}{\epsfxsize=1.4cm \epsffile{box6.eps}} &+&
  \quad {\cal O}(g^6).   \label{T1}
\end{eqnarray}
In the ladder approximation one only takes $V_1$ into account. 
Effectively, one then describes the full interaction between two nucleons by
\begin{eqnarray}
  \raisebox{-0.25cm}{\epsfysize=0.65cm\epsffile{T.eps}} =
  \raisebox{-0.25cm}{\epsfxsize=1.4cm \epsffile{oneba.eps}} &+&
  \raisebox{-0.25cm}{\epsfxsize=1.4cm \epsffile{onebb.eps}} \nonumber\\
+ \raisebox{-0.25cm}{\epsfxsize=1.4cm \epsffile{box2.eps}} +
  \raisebox{-0.25cm}{\epsfxsize=1.4cm \epsffile{box1.eps}} &+&
  \raisebox{-0.25cm}{\epsfxsize=1.4cm \epsffile{box3.eps}} +
  \raisebox{-0.25cm}{\epsfxsize=1.4cm \epsffile{box4.eps}} \nonumber\\
  &+& \quad {\cal O}(g^6). \label{T2}
\end{eqnarray}

In this approximation intermediate states containing more than three
particles do not occur.  This implies that time-ordered box diagrams
with four particles in the intermediate state are neglected, as we can
see if we compare the expansions in Eqs.~\r{T1} and ~\r{T2}.
As the individual diagrams contributing to $V_2$ are not covariant, the
sum of box diagrams produced by the ladder approximation is not
covariant.

Using equal-time quantization, twenty out of the twenty-four possible
time-orderings have intermediate states with more than three
particles.  On the LF, the spectrum condition kills many of the
time-ordered diagrams.  There are six nonvanishing diagrams, of which
four only contain two and three-particle intermediate states. One
concludes that the one-meson exchange kernel neglects the majority of the
contributing time-ordered box diagrams in equal-time quantization,
whereas on the LF most of the nonvanishing diagrams are taken
into account.  This does not mean necessarily that in IF dynamics the
ladder approximation misses most of the amplitude, since the missing
diagrams have smaller magnitudes.  The contribution of the missing
diagrams needs to be investigated in order to see how much the higher
Fock sectors are suppressed.

There is one thing which seems to complicate matters on the LF.  The
individual LF time-ordered diagrams are not rotational invariant.  When
a number of them is missing the full amplitude will also lack
rotational invariance, as is mentioned often in the literature.  This
feature does not occur on the equal-time plane, since here rotational
invariance is a manifest symmetry. However, in other types of
Hamiltonian dynamics other symmetries are nonmanifest.  In IF
Hamiltonian dynamics, e.g., boost invariance is not manifest.  
Therefore, we refer to breaking of covariance, which is a
general feature of any form of Hamiltonian dynamics, if one truncates the 
the Fock-space expansion.

We would like to estimate the contribution of the missing diagrams, 
irrespective of the strength of the coupling. 
It is not possible to do this in a completely general way, so we perform our
numerical calculations for the box diagram only. We assume that our results
will be indicative for the higher orders too.

We define the fraction
\begin{equation}
{\cal{R}}^{\rm LF}_{4^+} = 
\frac{\epsfxsize=1.4cm \epsffile{box0.eps} +
      \epsfxsize=1.4cm \epsffile{box6.eps}}
     {\raisebox{-.5cm}{\epsfxsize=1.4cm \epsffile{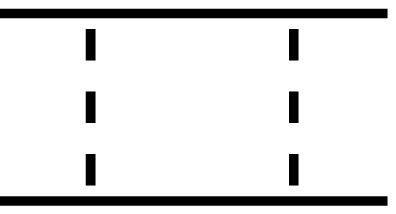}}}.
\end{equation}
The subscript 4 indicates that this variable includes all diagrams
having at least 4 particles in some intermediate state. 
For ${\cal{R}}^{\rm IF}_{4^+}$ one would have to add the diagrams 
containg five and six-particle intermediate states in the numerator,
as these give nonvanishing contributions in the instant-form.
The diagram in the denominator is the covariant diagram.

We shall show that the correction $V_2$ is indeed much less important
numerically in LFD than in IFD. 
So the 2N2m-state is
in LFD much less important than in IFD. We conjecture that this
property of LFD that the Fock-state expansion converges much more rapidly
than in IFD persists in higher orders in the coupling constant.

\section{The box diagram}
\label{secbox}
In the previous section we saw that the lowest level at which breaking
of covariance is to be expected is the two-meson exchange diagram, also
referred to as the box diagram. Implicitly, in Sec.~\ref{seclippmann}
we took the particles to be scalars, as we did not include instantaneous terms
which are related to spin-$1/2$ particles. Although a bound state of
scalar particles is not found in nature, we do not include spin because
we want to avoid in this investigation the complications due to 
instantaneous terms.

\begin{figure}
\hspace{2.5cm} \epsfxsize=2.9cm \epsffile{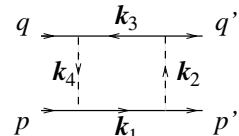}
\caption{\label{figkin}Kinematics for the box diagram. The arrows denote the momentum flow.}
\end{figure}

We look at the process of two nucleons with momenta $p$ and $q$
respectively, coming in and exchanging two meson of mass $\mu$. The
outgoing nucleons have momenta  $p'$ and $q'$. 
The kinematics is given in Fig.~\ref{figkin}.
The internal momenta are
\begin{eqnarray}
\label{eqk1}
k_1 &=& k,        \\
k_2 &=& k - p',   \\
k_3 &=& k - p - q,\\
k_4 &=& k - p.
\label{eqk4}
\end{eqnarray}
We have to keep in mind that these relations only hold for those components
of the momenta that are conserved.

\subsection{The covariant box diagram}
The covariant box diagram is given by
\begin{eqnarray}
\label{covbox}
\raisebox{-.3cm}{\epsfxsize=1.4cm \epsffile{boxcov.eps}}
= \int_{\rm Min} -i\;{\rm d}^4k \hspace{2cm}
\\
{\left( ({k_1}^2 - m^2) ({k_2}^2 - \mu^2)
        ({k_3}^2 - m^2) ({k_4}^2 - \mu^2) \right)^{-1}},
\nonumber
\end{eqnarray}
where the subscript ``Min'' denotes that the integration has a
Minkowskian measure. The imaginary parts $i \epsilon$ of the masses
have been dropped.

\subsection{Equivalence}

If the external states are on energy-shell, that is,
\begin{equation}
\label{onEcond}
P^- = p^- + q^- = p'^- + q'^-,
\end{equation}
then the time-ordered diagrams are the same as those derived
by integrating the covariant diagram over light-front energy $k^-$.
In that case we have
\begin{eqnarray}
\raisebox{-.3cm}{\epsfxsize=1.4cm \epsffile{boxcov.eps}}
=
\raisebox{-.25cm}{\epsfxsize=1.4cm \epsffile{box1.eps}}
+
\raisebox{-.25cm}{\epsfxsize=1.4cm \epsffile{box3.eps}}
+
\raisebox{-.25cm}{\epsfxsize=1.4cm \epsffile{box2.eps}}\nonumber\\
+
\raisebox{-.25cm}{\epsfxsize=1.4cm \epsffile{box4.eps}}
+
\raisebox{-.25cm}{\epsfxsize=1.4cm \epsffile{box0.eps}}
+
\raisebox{-.25cm}{\epsfxsize=1.4cm \epsffile{box6.eps}}
\label{equiv}
\end{eqnarray}
The example of the box diagram with scalar particles of equal mass
has been worked out before by Ligterink and Bakker~\cite{LB95b}.

\subsection{The LF time-ordered diagrams}

It is well-known~\cite{KS70,LB80} how to construct the LF time-ordered 
diagrams.  They are expressed in terms of integrals over energy denominators 
and phase-space factors. In the case of the box diagram we need the ingredients
given below.  The phase space factor is
\begin{equation}
\label{phasespace}
\phi = 16 | k_1^+ k_2^+ k_3^+ k_4^+ |.
\end{equation}
Without loss of generality we can take $p^+ \geq p'^+$. The internal
particles are on mass-shell, however, the intermediate
states are off energy-shell.
There are a number of possibilities for the intermediate states.
We label the corresponding kinetic energies according to which of the 
internal particles, labeled
by $k_1$ \dots $k_4$ in Fig.~\ref{figkin}, are in this state.
\begin{eqnarray}
\label{cut1a}
H_{14} &=& q^- + \frac{{k_1^\perp}^2 +  m^2}{2 k_1^+}
               - \frac{{k_4^\perp}^2 +\mu^2}{2 k_4^+}, \\
\label{cut2a}
H_{13} &=&       \frac{{k_1^\perp}^2 +  m^2}{2 k_1^+}
               - \frac{{k_3^\perp}^2 +  m^2}{2 k_3^+}, \\
\label{cut3a}
H_{12} &=& q'^-+ \frac{{k_1^\perp}^2 +  m^2}{2 k_1^+}
               - \frac{{k_2^\perp}^2 +\mu^2}{2 k_2^+}, \\
\label{cut1b}
H_{34} &=& p^- - \frac{{k_3^\perp}^2 +  m^2}{2 k_3^+}
               + \frac{{k_4^\perp}^2 +\mu^2}{2 k_4^+}, \\
\label{cut2b}
H_{24} &=&q'^-+p^-+\frac{{k_2^\perp}^2 +\mu^2}{2 k_2^+}
                  -\frac{{k_4^\perp}^2 +\mu^2}{2 k_4^+}, \\
\label{cut3b}
H_{23} &=& p'^-+ \frac{{k_2^\perp}^2 +\mu^2}{2 k_2^+}
               - \frac{{k_3^\perp}^2 +  m^2}{2 k_3^+}.
\end{eqnarray}
A minus sign occurs if the particle goes in the direction opposite to
the direction defined in Fig.~\ref{figkin}.
All particles are on mass-shell, including the external ones:
\begin{eqnarray}
q^- = \frac{{q^\perp}^2 +  m^2}{2 q^+},\hspace{1cm}
q'^- &=& \frac{{q'^\perp}^2 +  m^2}{2 q'^+},\nonumber\\
p^- = \frac{{p^\perp}^2 +  m^2}{2 p^+},\hspace{1cm}
p'^- &=& \frac{{p'^\perp}^2 +  m^2}{2 p'^+}.
\end{eqnarray}

We can now construct the LF time-ordered diagrams. The diagrams
\r{box1} and \r{box3} will be later referred to as trapezium diagrams,
\r{box2} as the diamond and \r{box0} as the stretched box.
%\begin{eqnarray}
%\label{box1}\raisebox{-0.25cm}{\epsfxsize=1.4cm \epsffile{box1.eps}} &=&
%\int {\rm d}^2k^\perp \int_{  0}^{p'^+} {\rm d}k^+
%\frac{-2 \pi}{\phi \; (P^-\!\!-H_{14}) \; 
                      %(P^-\!\!-H_{13}) \;(P^-\!\!-H_{12})},\\
%\label{box2}\raisebox{-0.25cm}{\epsfxsize=1.4cm \epsffile{box2.eps}} &=&
%\int {\rm d}^2k^\perp \int_{p'^+}^{p^+} {\rm d}k^+
%\frac{-2 \pi}{\phi \; (P^-\!\!-H_{14}) \; 
                      %(P^-\!\!-H_{13}) \;(P^-\!\!-H_{23})},\\
%\label{box3}\raisebox{-0.25cm}{\epsfxsize=1.4cm \epsffile{box3.eps}} &=&
%\int {\rm d}^2k^\perp \int_{p^+}^{p^++q^+} \hspace{-.5cm} {\rm d}k^+
%\frac{-2 \pi}{\phi \; (P^-\!\!-H_{34}) \; 
                      %(P^-\!\!-H_{13}) \;(P^-\!\!-H_{23})},\\
%\label{box0}\raisebox{-0.25cm}{\epsfxsize=1.4cm \epsffile{box0.eps}} &=&
%\int {\rm d}^2k^\perp \int_{p'^+}^{p^+} {\rm d}k^+
%\frac{-2 \pi}{\phi \; (P^-\!\!-H_{14}) \; 
                      %(P^-\!\!-H_{24}) \;(P^-\!\!-H_{23})},\\
%\raisebox{-0.25cm}{\epsfxsize=1.4cm \epsffile{box4.eps}} &=&
%\raisebox{-0.25cm}{\epsfxsize=1.4cm \epsffile{box6.eps}}= 0. \label{boxo}
%\end{eqnarray}
\begin{eqnarray}
\label{box1}\raisebox{-0.25cm}{\epsfxsize=1.4cm \epsffile{box1.eps}} \!\!&=&\!\!
\int {\rm d}^2k^\perp \int_{  0}^{p'^+} \!\!
\frac{-2 \pi \;{\rm d}k^+}{\phi \; (P^-\!\!-H_{14}) \; 
                      (P^-\!\!-H_{13}) \;(P^-\!\!-H_{12})},\\
\label{box2}\raisebox{-0.25cm}{\epsfxsize=1.4cm \epsffile{box2.eps}} \!\!&=&\!\!
\int {\rm d}^2k^\perp \int_{p'^+}^{p^+} \!\!
\frac{-2 \pi \;{\rm d}k^+}{\phi \; (P^-\!\!-H_{14}) \; 
                      (P^-\!\!-H_{13}) \;(P^-\!\!-H_{23})},\\
\label{box3}\raisebox{-0.25cm}{\epsfxsize=1.4cm \epsffile{box3.eps}} \!\!&=&\!\!
\int {\rm d}^2k^\perp \int_{p^+}^{p^++q^+} \hspace{-.5cm} \!\!
\frac{-2 \pi \;{\rm d}k^+}{\phi \; (P^-\!\!-H_{34}) \; 
                      (P^-\!\!-H_{13}) \;(P^-\!\!-H_{23})},\\
\label{box0}\raisebox{-0.25cm}{\epsfxsize=1.4cm \epsffile{box0.eps}} \!\!&=&\!\!
\int {\rm d}^2k^\perp \int_{p'^+}^{p^+} \!\!
\frac{-2 \pi \;{\rm d}k^+}{\phi \; (P^-\!\!-H_{14}) \; 
                      (P^-\!\!-H_{24}) \;(P^-\!\!-H_{23})},\\
\raisebox{-0.25cm}{\epsfxsize=1.4cm \epsffile{box4.eps}} \!\!&=&\!\!
\raisebox{-0.25cm}{\epsfxsize=1.4cm \epsffile{box6.eps}}= 0. \label{boxo}
\end{eqnarray}
The factor $2 \pi$ matches the conventional factor in Eq.~\r{covbox}.
The last two diagrams are zero because we have taken $p^+ \geq p'^+$ and
therefore these diagrams have an empty $k^+$-range. If we take
$p^+ \leq p'^+$, which case will also occur in upcoming sections, the diagrams \r{boxo}
have nonvanishing contributions.

\section{A numerical experiment}
\label{numexp}
 
\begin{figure}
\epsfxsize=8.8cm \epsffile{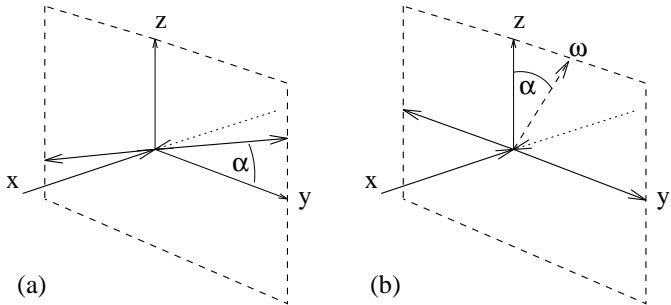}
\caption{(a) Two particles come in along the $x$-axis. They scatter
into the $y-z$ plane over an angle of $\pi/2$.  The azimuthal angle is
given by $\alpha$. (b) Another viewpoint.  The outgoing particles go
out along the y-axis. One can vary the orientation of the light-front
vector $\omega$ with respect to the $z$-axis.  \label{figscat}}
\end{figure}

We look at the scattering of two particles over an angle of $\pi/2$.
In Fig.~\ref{figscat} the process is viewed in two different ways.

Fig.~\ref{figscat}a pictures the situation where the scattering plane is
rotated around the $x$-axis.
The viewpoint in Fig.~\ref{figscat}b concentrates on
the influence of the orientation of the quantization plane, and is
connected to explicitly covariant LFD, as will be discussed in
Sec.~\ref{seconshell}.
Both viewpoints should render
identical results, since all angles between the five relevant directions
(the quantization axis and the four external particles) are the same.
We choose for the momenta
\begin{eqnarray}
\label{pmu}
p^\mu = ( v^0,&&+v^x ,\;\;\;\;\;   0,\;\;\;\;\;   0) ,\\
\label{qmu}
q^\mu = ( v^0,&&-v^x ,\;\;\;\;\;   0,\;\;\;\;\;   0),\\
\label{ppmu}
p'^\mu = ( v^0,&&\;\;\;\;\;  0 ,-v^y,-v^z),\\
\label{qpmu}
q'^\mu = ( v^0,&&\;\;\;\;\; 0 ,+v^y,+v^z).
\end{eqnarray}
indicating that we have chosen the fixed quantization plane $x^+=0$ 
(Fig.~\ref{figscat}a).
The incoming and outgoing particles have the same absolute values of
the momenta in the c.m. system. Therefore
\begin{equation}
|\vec{v}|^2 = (v^x)^2 = (v^y)^2 + (v^z)^2 = |\vec{v'}|^2.
\end{equation}
The Mandelstam variables are
\begin{eqnarray}
\label{mandels}
s &=& (p+q)^2 = 4 (v^0)^2, \\
\label{mandelt}
t &=& (p-p')^2 =-2 |\vec{v}|^2,\\
\label{mandelu}
u &=& (p-q')^2 = -2 |\vec{v}|^2.
\end{eqnarray}
We are now ready to perform the numerical experiments for three cases, 
which are described in Sec.~\ref{secversus}-\ref{secundert}. In the
experiments two parameters are focused on. We shall vary the 
azimuthal angle $\alpha$ in the $y$-$z$-plane,
\begin{equation}
\alpha = \arctan \frac{v^z}{v^y},
\end{equation}
and the incoming c.m.s.-momentum
\begin{equation}
v = v^x.
\end{equation}
In our experiments we will omit the units for the masses, which are MeV$/c^2$.

\section{Light-front dynamics versus instant-form dynamics}
\label{secversus}
One of the claims of LFD is that higher Fock states are more strongly
suppressed than in IFD. We can investigate this claim
for the box diagram easily in the following case.

We take the external states on energy-shell \r{onEcond}, such that the
equality \r{equiv} holds. At the same time we avoid on-shell
singularities for the intermediate states by giving the external
particles a slightly smaller mass $m'$,
\begin{equation}
\label{mprime}
{m'}^2 = p^2 < m^2,
\end{equation}
such that we can still relate the amplitude to a $S$-matrix element. 

The process we look at is described in the previous section  and
has two scalars of mass $m'$ coming in along
the $x$-axis, interacting, and scattered over a scattering angle of
$\pi/2$.  Stretched boxes give maximal contributions (see next section)
if the quantization axis is in the scattering plane, which is the case
if the azimuthal angle $\alpha = \pi/2$.

\begin{figure}
\hspace{1.5cm} \epsfxsize=5.5cm \epsffile{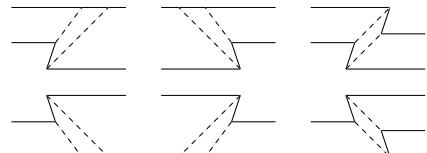}
\caption{\label{figdiagsR5} Time-ordered diagrams that contribute to
${\cal{R}}_5$. The diagrams in the first column have five
particles in the first intermediate state. The diagrams in the second
column have five particles in the last intermediate state, and the
diagrams on the right have five-particle intermediate states for both
the first and the third intermediate state.}
\end{figure}

${\cal{R}}^{\rm LF}_{4^+}$ is easily found by calculating the stretched
box.  ${\cal{R}}^{\rm IF}_{4^+} = {\cal{R}}^{\rm IF}_4 + 
{\cal{R}}^{\rm IF}_5 + {\cal{R}}^{\rm IF}_6$, however, has twenty nonzero
contributions. As an example, we show the six contributions to
${\cal{R}}_5$ in Fig.~\ref{figdiagsR5}. 
This helps us to understand why ${\cal{R}}^{\rm LF}_{5} = 0$. All
contributing diagrams contain vacuum creation or annihilation vertices,
which are forbidden by the spectrum condition.

There are twelve diagrams contributing to ${\cal{R}}_6$, and
all contain vacuum creation or annihilation vertices. Therefore
${\cal{R}}^{\rm LF}_{6}$ vanishes.

\begin{figure}
\epsfxsize=8.5cm \epsffile{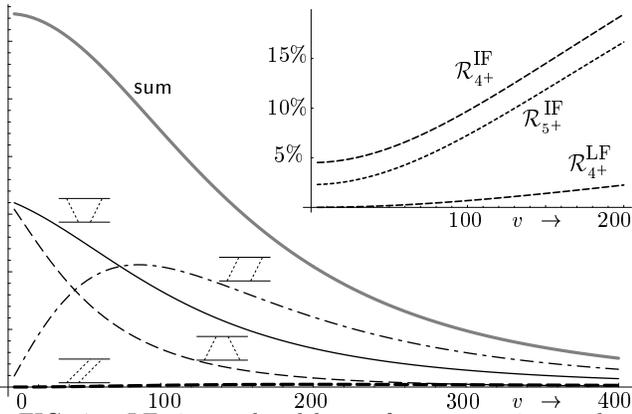}
\caption{\label{figR4and5} LF time-ordered boxes for a scattering angle
of $\pi/2$ as a function of the incoming momentum $v$. We also  
give the ratios of boxes with at least four particles 
(${\cal{R}}^{\rm IF}_{4^+}$ and ${\cal{R}}^{\rm LF}_{4^+}$) or five particles 
(${\cal{R}}^{\rm IF}_{5^+}$, ${\cal{R}}^{\rm LF}_{5^+} =0$) in one
of the intermediate states.} \end{figure}

We calculated
${\cal{R}}^{\rm IF}_{4^+}$ by subtracting the four diagrams only
containing three particle intermediate states from the full sum. This
sum can be obtained by doing the covariant calculation, or by adding
all LF time-ordered boxes.  Our results are given in
Fig.~\ref{figR4and5}. 
We also calculated ${\cal{R}}^{\rm IF}_{5^+}$.

We conclude that on the LF contributions of higher Fock states are
significantly smaller than in IFD. In the limit $v \rightarrow 0$ the
ratio ${\cal{R}}^{\rm LF}_{4^+}$ goes to zero, because the phase space
becomes empty. However, in IFD there is a finite contribution of
${\cal{R}}^{\rm IF}_{4^+} = 4.5 \%$ in this limit.  Even if one
includes five-particle intermediate states the LF is the winner by far.

Note that $m'$, given by Eq.~\r{mprime}, varies as a function of $p^2$,
and therefore also as a function of $v$, but is independent of
$\alpha$. The deviation of $m'$ from $m$ is small; less than 2.3\% for
$v < 200$, and less than 9\% for $v < 400$.  As the deviation of the
mass $m'$ from $m$ is only small, we are convinced that these results are
indicative for calculations above threshold. However, we do not want to
do these calculations, because then one needs to subtract the on-shell
singularities of the equal-time ordered boxes.

\section{Numerical results above threshold}
\label{secabovet}

As in Sec.~\ref{secversus} we look at the scattering of two particles
over an angle of $\pi/2$. We focus on LFD, and therefore we simply
write ${\cal{R}}_4 = {\cal{R}}^{\rm LF}_{4^+}$.  We do no try to avoid
on-shell singularities by taking different masses for the internal and
external nucleons.  Two nucleons of mass $m= 940$  scatter via the
exchange of scalar mesons of mass $\mu = 140$.  Again, the vertex is a
scalar and no spin is included.

\subsection{Evaluation method}
Contrary to the case considered in Sec.~\ref{secversus}, 
now it is not straightforward
to evaluate the contributions of the LF time-ordered boxes, because
the nonstretched boxes contain on-shell singularities, thoroughly
analyzed in Sec.~\ref{seconshell}. Here we briefly sketch
how we deal numerically with the singularities. Using the analysis
of Sec.~\ref{seconshell} we identify the singularity $I_{\rm sing}$, 
and rewrite the nonstretched boxes as
\begin{equation}
\label{singsub}
\int {\rm d}^3k \; I = \int {\rm d}^3k (I - I_{\rm sing})
+ \int {\rm d}^3k \; I_{\rm sing}.
\end{equation} 
The integrand $I_{\rm sing}$ has a simple algebraic form, such that
the integration in one dimension over the singularity can be done analytically,
and the remaining integral is regular. This integral is then done
numerically by {\sc mathematica}. 
The integral over $(I - I_{\rm sing})$ was implemented in {\sc fortran}. 
These two numbers are added to give the results presented is
Secs.~\ref{subsalpha} and \ref{subsvx}.

\subsection{\label{subsalpha}Results as a function of $\alpha$}
We shall now vary the direction of $\vec{v'}$, given by the azimuthal angle
$\alpha$, however, not its length. Therefore the Mandelstam variables are
independent of $\alpha$, and we expect the full amplitude to be invariant.
We tested this numerically for a number of values of ${v}$.
In the region $0 \leq \alpha \leq \pi$ we used the formulas \r{box1} until
\r{box0}. In the region $\pi \leq \alpha \leq 2 \pi$ the diagrams
\r{box2} and \r{box0} vanish. However, then there are contributions
from the diagrams in \r{boxo}.
The results are shown in Fig.~\ref{figabove}.
\def \figurefive{
\begin{figure*}
\epsfxsize=18cm \epsffile{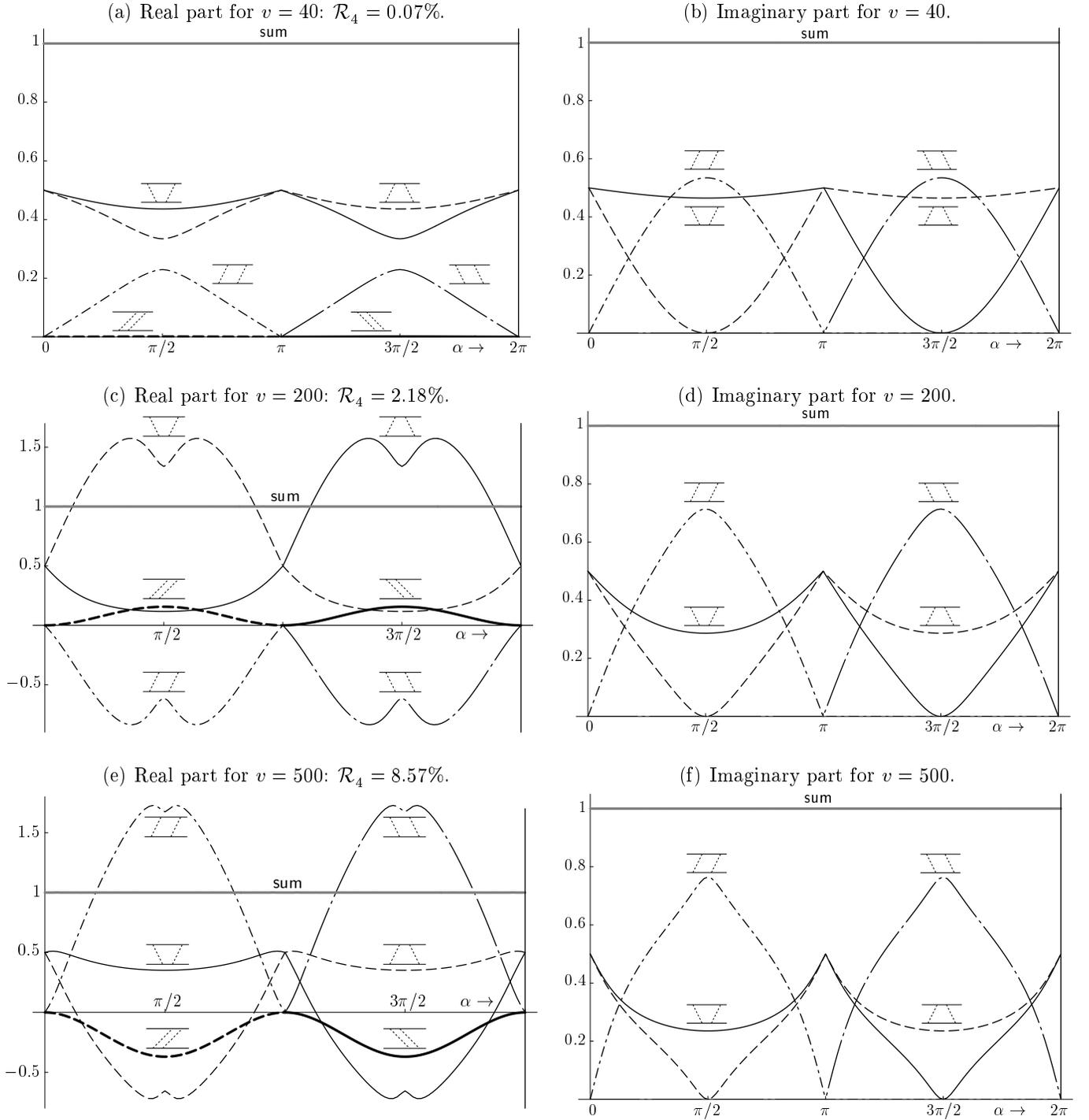}
\caption{\label{figabove}Amplitudes above threshold from $\alpha = 0$ to $\alpha = 2 \pi$.
${\cal{R}}_4$ is the maximal fraction of the stretched box to the absolute value
of the sum.}
\end{figure*}
}
The results are normalized to the value of the covariant amplitude.
The contributions from the different diagrams vary strongly with the
angle $\alpha$. Since the imaginary parts are always positive, they are
necessarily in the range $[0,1]$ when divided by the imaginary part of
the covariant amplitude. The real parts can behave much more excentric,
especially for higher values of the incoming c.m.s.-momentum $v$. An
analysis of the $\alpha$-dependence is given in Sec.~\ref{seconshell}.
Clearly the LF time-ordered diagrams add up to the covariant amplitude,
so we see that in all cases we obtain covariant (in particular
rotationally invariant) results for both the real and the imaginary
part.

\subsection{\label{subsvx}Numerical results as function~of~${v}$}
We look at scattering in the $x$-$z$-plane ($\alpha = \pi/2$),
because in that case the contributions from the
stretched boxes are maximized. The results are shown in Fig.~\ref{figsix}.
\def \figuresix{
\begin{figure*}
\epsfxsize=18cm \epsffile{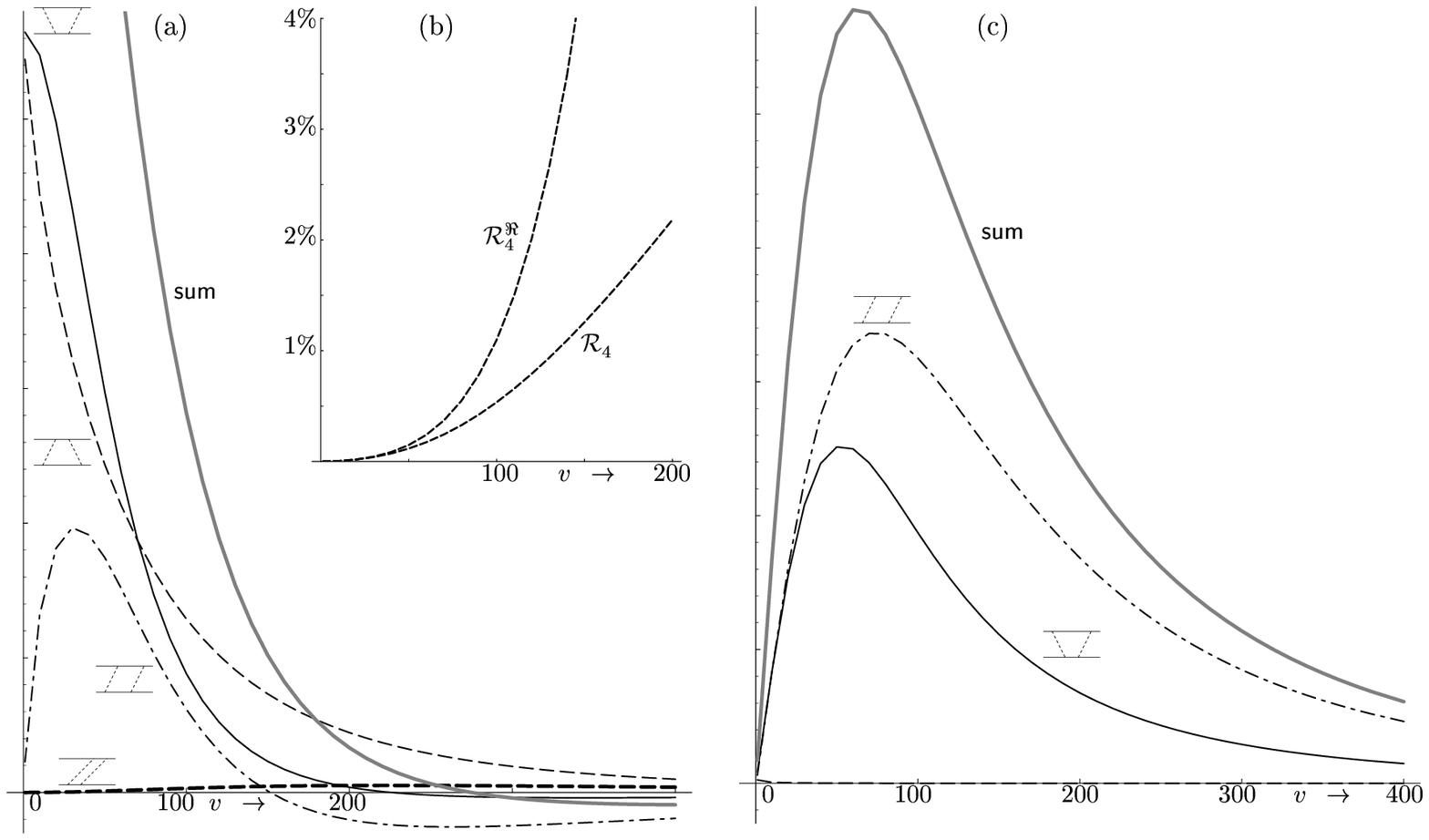} 
\caption{\label{figsix} 
Real (a) and imaginary part (c) of the LF time-ordered boxes above threshold
for $\alpha = \pi/2$ as a function of the momentum
of the incoming particles $v$. The inset (b) shows the ratio of
the stretched box to the real part of the amplitude (${\cal R}_4^{\Re}$)
and to the absolute value (${\cal R}_4$).}
\end{figure*}
}

We see that the stretched box, the diagram with two
simultaneously exchanged mesons, is relatively small at low energies, but becomes
rather important at higher energies. 
We depict the ratio of the stretched box to both the real part and to the
total amplitude. Since the real part has a zero near $v= 280$, the
first ratio becomes infinite at that value of the incoming momentum. 

\twocolumn[\hsize\textwidth\columnwidth\hsize\csname
@twocolumnfalse\endcsname
\figurefive
]

\twocolumn[\hsize\textwidth\columnwidth\hsize\csname
@twocolumnfalse\endcsname
\figuresix
]

\section{Numerical results off energy-shell}
\label{secundert}

In the previous section, we tested covariance of the LF formalism for
amplitudes with on energy-shell external particles, 
by using the c.m. frame, where $P^\perp = p'^\perp + q'^\perp = 0$
and $P^z = p'^z + q'^z = 0$.
 However, on the LF the operator $P^z$ is  dynamical,
 and the last equality does not hold anymore off energy-shell,
as one can easily verify in the following way. 
Consider the case of a bound state with mass ${\cal M} < 2m$.
The bound state is off energy-shell, and therefore we have
\begin{equation}
\label{eq1653}
P^- < p'^- + q'^-.
\end{equation} 
The plus and transverse momenta are kinematic, so
\begin{eqnarray}
\label{eq1253}
P^+     &=& p'^+ + q'^+, \\
\label{eq1654}
P^\perp &=& p'^\perp + q'^\perp.
\end{eqnarray} 
Adding Eqs.~\r{eq1653} and \r{eq1253} gives
\begin{equation}
\label{eq1655}
P^z < p'^z + q'^z.
\end{equation} 
If $P^z = 0$, then Eq.~\r{eq1655} implies that $p'^z + q'^z > 0$.
Therefore  the two outgoing particles cannot have exactly opposite momenta
as in Eqs.~\r{ppmu} and \r{qpmu}. In terms of the explicitly covariant
LFD, introduced in Sec.~\ref{seconshell}, this reflects the fact that the off  
energy-shell relation between $p'+q'$ and $P$ contains extra four-momentum
like in Eq.~\r{eq29}.
What was the reason that
we chose opposite momenta in the previous sections in the first place? 
Our reason  was that
we wanted to have a manifest symmetry of the amplitude, because it
is obvious that the Mandelstam variables $s$, $t$ and $u$ remain
the same under the rotations we investigated. We investigate
how the off-shell amplitude breaks covariance by using momenta 
for the outgoing particles, such that the Mandelstam variables remain constant,
at the same time satisfying the conditions 
\r{eq1653}-\r{eq1655}.

As in the previous sections, we shall fix the direction of the incoming
particles, as in Eqs.~\r{pmu} and ~\r{qmu}, and vary the direction of
the outgoing particles. Therefore, by construction, $s$ is invariant.
%The only parts of the expression for the Mandelstam variables 
%$t$ \r{mandelt} and $u$ \r{mandelu} which are not
%manifestly invariant are $\inp{p}{q'}$ and $\inp{p}{p'}$.  These can be
%written as
In order to guarantee the invariance of the other Mandelstam variables
$t$ and $u$ we must take $\inp{p}{q'}$ and $\inp{p}{p'}$ constant, even
off energy-shell. These inner products are:
%We will write these inner products in terms of the
%components of $p'$ and $q'$:
\begin{eqnarray} \label{inpq}
 \inp{p}{q'} = p^+ q'^- + q'^+ p^-  -
 \inp{p^\perp}{q'^\perp} \hspace{30mm} \\
 =x_p P^+\frac{{q'^\perp}^2\!\!+ m^2}{2 (1\!-\!x_{p'}) P^+} +
 (1\!-\!x_{p'}) P^+ \frac{{p^\perp}^2\!\!+ m^2}{2 x_p P^+} - \inperp{p}{q'} ,
 \hspace{-1cm} \nonumber \\ 
 \label{inpp}
 \inp{p}{p'} = p^+ p'^- + p'^+ p^- - \inperp{p}{p'} \hspace{30mm} \\
 = x_p P^+ \frac{{q'^\perp}^2 + m^2}{2 x_{p'} P^+} + x_{p'} P^+
	  \frac{{p^\perp}^2 + m^2}{2 x_p P^+}  - \inperp{p}{p'}.
 \hspace{-1cm} \nonumber 
\end{eqnarray} 
We have introduced the fractions 
\begin{equation}
\label{xfrac}
x_p = p^+/P^+, \quad x_{p'}=p'^+/P^+.
\end{equation}
Since the perpendicular momenta are conserved we have in the c.m. system
$p'^\perp = - q'^\perp, $
so the inner products of the
perpendicular momenta can be written as 
\begin{eqnarray}
\label{inperppp} \inperp{p}{p'} = \phantom{-} |p^\perp| |p'^\perp| \cos \theta,  
\\ 
\label{inperppq} \inperp{p}{q'} = - |p^\perp| |p'^\perp| \cos \theta, 
\end{eqnarray} 
where $\theta$ is the scattering angle.  We can now solve 
Eqs.~(\ref{inpq}-\ref{inperppp}) for $x_{p'}$, $|p'^\perp|$ and
$\theta$. There are many curves satisfying these conditions. For uniqueness we demand
that the curve goes through the point in which $x_{p'} = x_p = 1/2$, $|p'^\perp|
= |p^\perp|$ and $\theta = \pi/2$.  We find that the curve is then parametrized by

\begin{eqnarray} 
\label{condtheta} \theta &=& \pi/2 ,\\ 
\label{condperp}
\frac{{p'^\perp}^2 + m^2}{x_{p'} (1\!-\!x_{p'})} &=& 4 ({p^\perp}^2 + m^2).
\end{eqnarray}

From $x_{p'}$ and $p'^\perp$ we determine $p'^+$ and $p'^-$ , and
finally also $p'^z$. Because the particles come in along the $x$-axis, the
above relations define an ellipse in the $y$-$z$-plane. In the case of IFD
these ellipses would reduce to circles with center of the origin and radius $v$.
\begin{figure} 
\epsfxsize=8cm \epsffile{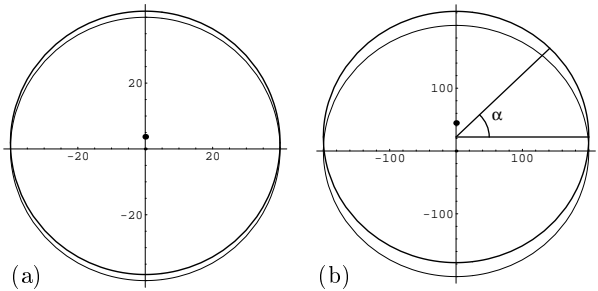}
\caption{\label{rotinv} Possible momenta $p'$ and $q'$ of the outgoing
particles  (fat line) in the scattering plane (horizontal: $p'^y$ and $q'^y$,
vertical: $p'^z$ and $q'^z$) for two cases: (a) $v=40$
and (b) $v = 200$. The momentum $p'+ q'$ is indicated by the dot. As a
reference we have drawn the locus for on energy-shell external
particles (thin line): a circle centered at the origin. }
%The angle $\alpha$ is indicated in (b).}
\end{figure} 
In Fig.~\ref{rotinv} we have
indicated the $y$ and $z$-components of the momenta of the external
particles for the two cases we investigate.  We see that
Eqs.~\r{eq1654} and \r{eq1655} hold.  The off
energy-shell momenta form an ellipse. The deviation from a circle with
radius $v$ is hardly visible.  

\begin{figure}
\epsfxsize=8.5 cm \epsffile{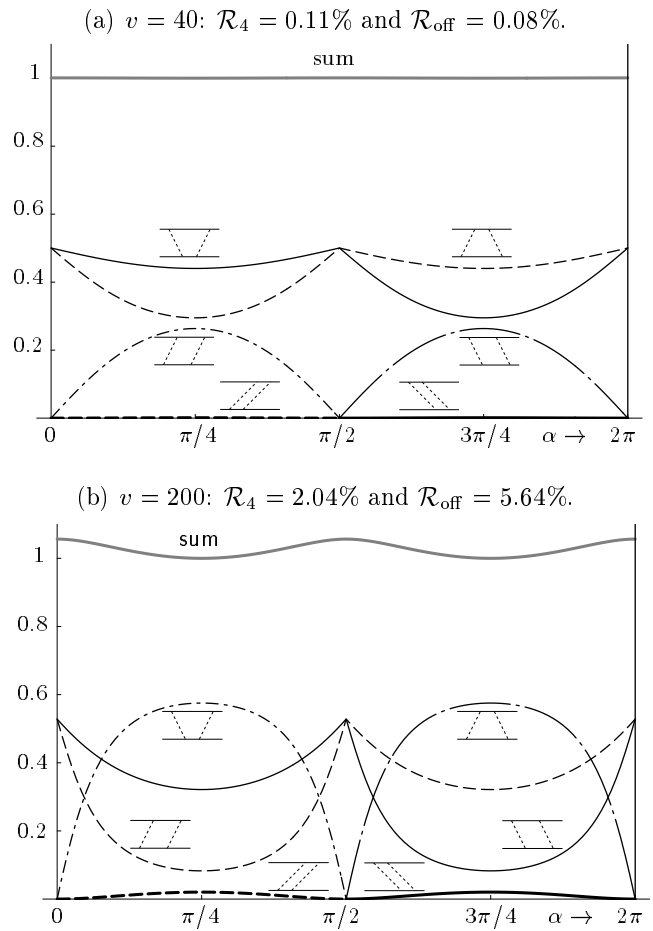}
\caption{\label{figoff} The LF time-ordered boxes as a function of
$\alpha$.}
\end{figure}

In Fig.~\ref{figoff} we show the
contributions of the different boxes and their sum as we vary the angle
$\alpha$. The calculations are the same as in the previous section,
using the formulas \r{box1}-\r{boxo}, except that the momenta of the
outgoing particles have changed such that \r{condtheta} and
\r{condperp} are satisfied. As there does not exist a covariant amplitude in
the off energy-shell case, we normalized the curves shown by dividing the amplitudes
by their sum at $\alpha = \pi/4$.

\begin{figure}
\epsfxsize=8.5cm \epsffile{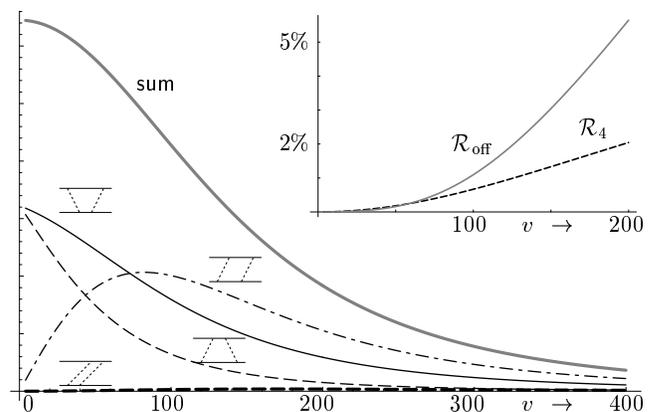}
\caption{\label{figmaxoff} The LF time-ordered boxes as a function of
$v$ for $\alpha = \pi/2$. The inset shows the
maximum contributions of the stretched box, and the maximal breaking
of covariance. }
\end{figure}

In Fig.~\ref{figoff} we see the consequences of the off energy-shell
initial and final states. Condition~\r{onEcond} is violated, therefore
Eq.~\r{equiv} does not hold and a breaking of covariance can be expected.  
We see that the contributions of the higher Fock states ${\cal R}_{4}$
are smaller than the effect of the off-shellness ${\cal R}_{\rm off}$.
This is confirmed in Fig.~\ref{figmaxoff}, in which $\alpha$ is fixed 
and the incoming c.m.s-momentum $v$ is varied. 

From Fig.~\ref{figoff} we infer that the full amplitude is maximal at
$\alpha = 0$  and $\alpha= \pi$. The minimum is reached at 
$\alpha = \pi/2$ and $\alpha= 3 \pi/2$. Therefore the maximal breaking
of covariance of the amplitude can be calculated by taking the
difference of the total sum at the angles $\alpha = 0$ and
$\alpha = \pi/2$. We see that at typical values for incoming
momentum ($v \leq 50$) ${\cal R}_{\rm off}$ is small, even smaller
than ${\cal R}_{4}$. However, at higher momenta it 
dominates over the stretched box. In this region we see that 
the stretched box contributions remain small. 

A detailed explanation of the results for off energy-shell amplitudes is given 
in Sec.~\ref{secoffshell}.

\section{Analysis of the on energy-shell results}
\label{seconshell}
The angle dependence of the LF time-ordered amplitudes found numerically,
can be understood analytically. 
The variation of the LF amplitudes with the 
angle $\alpha$ means that they have singularities in this variable, either
at finite values of $\alpha$ or at infinity. 
They should disappear when they are summed to give the covariant amplitude. 
These singularities can be 
most conveniently analyzed in the explicitly covariant version of LFD 
(see for a review \cite{CDKM98}). In this version the orientation of the 
light-front is given by the invariant equation 
$\omega\cd x=0$. The amplitudes are calculated by the rules of the graph 
technique explained in Ref.~\cite{CDKM98}. 
After a transformation of variables, these amplitudes
coincide with those given by ordinary LFD. However, they are 
parametrized in a different way. The dependence of the amplitudes on 
the angle $\alpha$ means, in the covariant version, that they depend on the 
four-vector $\omega$ determining the orientation of the LF plane:  
$M=M(p,q,p',q',\omega)$. Hence, besides the usual Mandelstam variables 
$s$ \r{mandels} and 
$t$ \r{mandelt} the amplitude $M$ depends on the scalar products of $\omega$ 
with the four-momenta.  Since $\omega$ determines the direction only 
(the theory is invariant relative to the substitution 
$\omega\rightarrow a\omega$), an amplitude should depend on the 
ratios of the scalar products of the four-momenta with $\omega$. 
Hence~\cite{Kar78}:
\begin{equation}\label{eq1a}
M=M(s,t,x_p,x_{p'}),
\end{equation}
where
\begin{equation}\label{eq2a}
x_p=\frac{\omega\cd p}{\omega\cd (p+q)},\quad
x_{p'}=\frac{\omega\cd p'}{\omega\cd (p+q)}.
\end{equation}
The formulas \r{eq2a} coincide with the definitions \r{xfrac} if we use
the $z$-axis as the quantization axis.
The $\omega$-dependence is reduced to two scalar variables 
$x_p$ and $x_{p'}$, since the direction of $\vec{\omega}$ is determined by two angles.
Hence, this amplitude should have singularities in the variables $x_p$ and $x_{p'}$.
Their positions will be found below.  The amplitude corresponding to 
the sum of all time-ordered diagrams should not depend on 
$x_p$ and $x_{p'}$. 

\begin{figure}
\hspace{1.1cm} \epsfxsize=6cm \epsffile{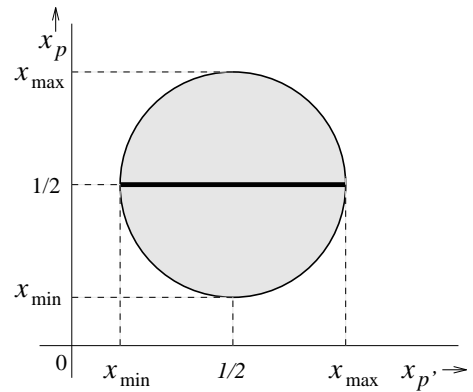}
\caption{\label{figphysreg}Physical region in the $x_p$-$x_{p'}$-plane.}
\end{figure}

Let us find the physical domain of the variables $x_p$ and $x_{p'}$, 
corresponding to all possible directions of $\vec{\omega}$ for fixed 
$s,t$. In the c.m. system the variables Eq.~(\ref{eq2a}) are represented 
as:
\begin{equation}\label{eq3b}
x_p=\frac{1}{2}-\frac{v}{\sqrt{s}}
\inp{\hat{\omega}}{ \hat{p}}, \quad
x_{p'}=\frac{1}{2}-\frac{v}{\sqrt{s}}
\inp{\hat{\omega}}{ \hat{p}'},
\end{equation}
where, e.g., $\inp{\hat{\omega}}{ \hat{p}}$ is the scalar product of the unit
vectors $\hat{\omega} = \vec{\omega}/|\vec{\omega}|$ and $\hat{p} = \vec{p}/|\vec{p}|$
in three-dimensional Euclidian space, and $v=\sqrt{s/4-m^2}$ is the
momentum of the particle in the c.m. system.  The Eqs.~(\ref{eq3b})
determine an ellipse in the $x_p$-$x_{p'}$-plane.  Its boundary is
obtained when $\vec{\omega}$ is in the scattering plane, that is
$\inp{\hat{n}}{ \hat{p}} = \cos \gamma$ or $\inp{\hat{n}}{ \hat{p}} = \cos (\gamma -
\theta)$, where $\gamma$ is the angle between  $\vec{p}$ and
$\vec{\omega}$ in coplanar kinematics and $\theta$ is the
scattering angle in the c.m.s. The case when $\vec{\omega}$ is out of
the scattering plane corresponds to the interior of the ellipse. For a
scattering angle $\theta=\pi/2$ the ellipse turns into a circle,
shown in Fig.~\ref{figphysreg}.

For the kinematics shown in Fig.~\ref{figscat} and Eqs.~\r{pmu}-\r{qpmu},
i.e., when $\vec{\omega}\perp \vec{p}$, it follows from Eq.~(\ref{eq3b}) 
that the value $x_p$ is fixed: $ x_p=\frac{1}{2},  $
whereas for a given $\alpha$ we obtain:
\begin{equation}\label{eq3d}
x_{p'}=\frac{1}{2}-\frac{v}{2v_{0}}\sin\alpha,
\end{equation}
with $v_{0}=\sqrt{m^{2}+v^{2}}$,
varies along a straight line when $\hat{\omega}$ is rotated in the $y$-$z$-plane. 
The bounds of the physical region of $x_{p'}$ are:
\begin{equation}\label{eq3c} 
 x_{\rm min} = \frac{1}{2}-\frac{v}{2v_{0}},\quad 
x_{\rm max} = \frac{1}{2}+\frac{v}{2v_{0}}. 
\end{equation}
When $0\leq \alpha \leq \pi/2$, $x_{p'}$ moves from 1/2 to $x_{\rm min}$.
When $ \pi/2 \leq \alpha \leq  \pi$, $x_{p'}$ moves in the 
opposite direction in the same interval.
This explains why all the curves in Figs.~\ref{figabove} 
and~\ref{figoff}  in the interval $0\leq \alpha \leq\pi$ are symmetric
relative to $\alpha=\pi/2$.

When $\pi \leq \alpha \leq 3\pi/2$, $x_{p'}$ moves from 1/2 to $x_{\rm
max}$ and, finally, when $3\pi/2 \leq \alpha \leq 2\pi$, it goes back
in the same interval. As in the previous paragraph, this explains why
all the curves in Figs.~\ref{figabove} and ~\ref{figoff} in the
interval $\pi\leq \alpha\leq 2\pi$ are symmetric relative to
$\alpha=3\pi/2$.
When $\alpha=\pi/2$ and $3\pi/2$, the values of $x_{p'}$ 
are on the boundary of the physical region.

Note also that the amplitudes for the trapezium (dashed and solid curves in 
Figs.~\ref{figabove} and~\ref{figoff}) 
are evidently obtained by the replacement 
$p\leftrightarrow  q, \quad p'\leftrightarrow  q'$, which, according to 
the definition in Eq.~(\ref{eq2a}), corresponds to $x_{p'}\rightarrow 
1-x_{p'}$. This is the same as the
replacement $\alpha\rightarrow 2\pi-\alpha$ in Eq.~(\ref{eq3d}). 
Therefore the curves for the other trapezium, when $\alpha$ goes 
from $2\pi$ to 0, repeat the curves for the trapezium, when $\alpha$ 
increases from 0 to $2\pi$. The same is true for the other diagrams 
(diamonds and stretched boxes).

%%%%%%%%%%%%%%%%%%%%%%%%%%%%%%%%%%%%%%%%%%%%%%%%%%%%%%%%%
\subsection{Trapezium}\label{trap}
\begin{figure}
\hspace{2.5cm} \epsfxsize=2.9cm \epsffile{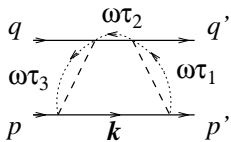}
\caption{\label{figtrap}The trapezium in explicitly covariant LFD.}
\end{figure}

The method of finding the singularities of the LF diagrams was 
developed in~\cite{Kar78}. Here we restrict ourselves to the example of   
the diagram~\r{box1}. Its counterpart in the explicitly
covariant LFD is shown in Fig.~\ref{figtrap}.
The dotted lines in this figure are associated with fictitous
particles (spurions), with four-momenta proportional to $\omega$. 
The four-momenta of the particles (the spurions not included) are not 
conserved in the vertices. Conservation is restored
by taking into account the spurion four-momentum. In the ordinary 
LF approach this corresponds to nonconservation of the 
minus-components of the particles. The spurions make up for the difference.

According to the 
rules of the graph techniques~\cite{CDKM98}, one should associate with 
a particle line with four-momentum $p$ and mass $m$  the factor 
$\theta(\omega\cd p)\delta(p^2-m^2)$ and associate with the spurion line with
four-momentum $\omega\tau_i$ the factor $1/(\tau_i-i\epsilon)$. 
Then one integrates, with measure d$^4k_i/(2\pi)^3$, over all the 
four-momenta $k_i$ not 
restricted by the conservation laws in the vertices, and over all 
${\rm d}\tau_i$.  The expression for the amplitude of Fig.~\ref{figtrap} is:
\begin{eqnarray}\label{eq1} 
\Mone&=&\int\theta(\omega\cd 
k)\delta(k^2-m^2) \nonumber \\ &&\times\theta(\omega\cd (p'-k))\ 
\delta((p'-k + \omega\tau_1)^2-\mu^2)\nonumber \\ 
&&\times\theta(\omega\cd (p+q-k))\ \delta((p+q-k+\omega\tau_2)^2-m^2) 
\nonumber \\ &&\times\theta(\omega\cd (p-k))\ 
\delta((p-k+\omega\tau_3)^2-\mu^2) \nonumber \\ 
&&\times\frac{{\rm d}\tau_1}{(\tau_1-i\epsilon)}\frac{{\rm d}\tau_2}{(\tau_2-i\epsilon)} 
\frac{{\rm d}\tau_3}{(\tau_3-i\epsilon)}\frac{{\rm d}^4k}{(2\pi)^3}\ .  
\end{eqnarray}                                                                    
Like in Eq.~\r{covbox}, we omit the coupling constant.
Performing the integrals over $\tau_i$ and ${\rm d}k_{0}$ in Eq.~(\ref{eq1})
by means of the $\delta$-functions, we get:
\begin{eqnarray}\label{eq1p}
\Mone=\int\frac{\theta(\omega\cd (p'-k))}{\mu^2-(p'-k)^2}\ 
\frac{\theta(\omega\cd (p+q-k))}{m^2-(p+q-k)^2} \nonumber\\
\label{eq2} \times \; \frac{\theta(\omega\cd (p-k))}{\mu^2-(p-k)^2}\ 
\theta(\omega\cd k)\frac{{\rm d}^3k}{2\varepsilon_k(2\pi)^3}.
\end{eqnarray}                                                                    

By transformations of variables (see for the details appendix B 
of~\cite{CDKM98}) the expression Eq.~(\ref{eq2}) can be transformed such that it
exactly coincides with Eq.~(\ref{box1}). 

For Feynman amplitudes the method to find their singularities was developed 
by Landau in \cite{Lan59}. A method very similar to that one can be applied to
time-ordered amplitudes.
If we would omit for a moment the $\theta$-functions in Eq.~(\ref{eq2}) 
and would not take into account that $k^2=m^2$, for finding the singularities
we should construct the function 
$\varphi_1=\alpha_1(\mu^2-(p'-k)^2)+\alpha_2(m^2-(p+q-k)^2) 
+\alpha_3(m^2-(p-k)^2)$ formed from the denominator of (\ref{eq1p}).
The singularities of the trapezium are found by putting to zero the 
derivatives of $\varphi_1$ with respect to $\alpha_{1-3}$ and 
with respect to $k$.  However, the trapezium may have singularities 
corresponding to a coincidence of the singularities of its integrand 
with the boundary of the integration domain caused by the presence of the
$\theta$-functions. 
So, we must find a conditional extremum. The restrictions 
can be taken into account using Lagrange multipliers \cite{Kar78}. 
Hence,  we should consider the function:
\begin{eqnarray}
\varphi=\alpha_1(\mu^2-(p'-k)^2)+\alpha_2(m^2-(p+q-k)^2) \nonumber\\
\label{eq3}
+\alpha_3(\mu^2-(p-k)^2)+\alpha_4(k^2-m^2)
+\gamma_1 \omega\cd (k-p'),
\end{eqnarray}
where $\alpha_4$ and $\gamma_1$ are the Lagrange multipliers.
One should also consider the functions obtained from 
$\varphi$ at $\alpha_1=0$, then at $\alpha_2=0$, then $\alpha_3=0$, 
then  at $\alpha_1=\alpha_2=0$, etc.  One should not consider the 
function obtained from (\ref{eq3}) by $\alpha_4=0$, since the 
integral (\ref{eq2}) contains the three-dimensional integration volume
${\rm d}^3k$, and the condition $k^2=m^2$ cannot be deleted. 
Therefore there is no need to introduce the term 
$\gamma_2\omega\cd k$, since the $k^2=m^2$ condition prevents
this term from being zero and, hence, does not impose any restrictions. 
The case $\gamma_1=0$ reproduces the singularities of the Feynman 
graph. Therefore below we shall consider the case 
$\gamma_1\neq 0$ resulting in the singularities in the 
variables $x_p,x_{p'}$.
We suppose that  
$\omega\cd p \geq \omega \cd p'.$ 
This corresponds to the condition $p^+\geq p'^{+}$ of Sec.~\ref{secbox}.
In the kinematics shown in Fig.~\ref{figscat} this means that $x_{p'}\leq
1/2$ and $0\leq \alpha\leq \pi$. In this case, the second and the third
$\theta$-functions in Eq.~(\ref{eq2}) do not give any restrictions and can be
omitted. Therefore we omit also the term $\gamma_3 \omega\cd (p-k) +
\gamma_4 \omega\cd (p+q-k)$.

The derivatives of $\varphi$ with respect to $k$, the $\alpha$'s and $\gamma_1$ give:
\begin{eqnarray}
\partial \varphi/\partial k =\alpha_1 2(p'-k)+\alpha_2 2(p+q-k) \nonumber\\
\label{eq19}   +\alpha_3 2(p-k)+\alpha_4 2k+\gamma_1 \omega=0,
\end{eqnarray}
with
\begin{eqnarray}
(p'-k)^2 & = & \mu^2, \quad (p+q-k)^2=m^2, \nonumber\\
\label{eq18a}
(p-k)^2  & = & \mu^2, \quad k^2  =  m^2, \quad \omega\cd k=\omega\cd p'.
\end{eqnarray}

\twocolumn[\hsize\textwidth\columnwidth\hsize\csname
@twocolumnfalse\endcsname
We multiply Eq.~(\ref{eq19}) in turn by $(p'-k)$, $(p+q-k)$, et cetera, and get:
\begin{equation}\label{eq19a}
\begin{array}{ll}
(\ref{eq19})\times p'-k:&  \alpha_1 2\mu^2 +\alpha_2 \mu^2
 +\alpha_3(2\mu^2-t) -\alpha_4 \mu^2 =0,\\ 
(\ref{eq19})\times p+q-k:& \alpha_1\mu^2 +\alpha_2 2m^2  
+\alpha_3 \mu^2 +\alpha_4 (s-2 m^2) +\gamma_1 (1-x_{p'})=0,\\ 
(\ref{eq19})\times p-k:& \alpha_1 (2\mu^2-t)+\alpha_2 \mu^2 
+\alpha_3 2\mu^2-\alpha_4 \mu^2 +\gamma_1 (x_p-x_{p'}) =0,\\
(\ref{eq19})\times k:&-\alpha_1\mu^2 +\alpha_2 (s-2m^2) 
-\alpha_3 \mu^2+2\alpha_4 m^2 + \gamma_1 x_{p'} =0, \\ 
(\ref{eq19})\times \omega: & \alpha_2 (1-x_{p'}) +\alpha_3 
(x_p-x_{p'}) 
+\alpha_4 x_{p'}=0 .
\end{array} 
\end{equation}
These equations have a nontrivial solution iff:
\begin{equation}\label{eq19b}
\left|
\begin{array}{lllll}
2\mu^2       & \mu^2   &(2\mu^2-t)   & -\mu^2   & 0 \\ 
\mu^2        & 2m^2    & \mu^2       &(s-2 m^2) &(1-x_{p'})\\ 
(2\mu^2-t)   & \mu^2   & 2\mu^2      &-\mu^2    &(x_p-x_{p'}) \\
-\mu^2       & (s-2m^2)& -\mu^2      & 2m^2     &  x_{p'}  \\ 
0            & (1-x_{p'}) & (x_p-x_{p'})   & x_{p'}      & 0  
\end{array} 
\right| =0.
\end{equation}
]
Eq.~(\ref{eq19b}) is quadratic in $x_{p'}$. Its 
solution is simple but lengthy. We show it for the particular case of 
the kinematics of Fig.~\ref{figscat} supposing that the particles in the
c.m. system have momenta~$v$. In this case 
$s$ and $t$ are given by Eqs.~(\ref{mandels}) and (\ref{mandelt}).
The solution of Eq.~(\ref{eq19b}) is:
\begin{equation}\label{eq19d}
x_{p'}^0=\frac{1}{2}\pm \frac{v}{2v_{0}}
\frac{\sqrt{2\mu^4+8\mu^2v^2+4v^4}}
{\sqrt{\mu^4+8\mu^2v^2+4v^4}}.
\end{equation}
The position of singularity in the variable $x_{p'}$ is denoted by $x^0_{p'}$.  
According to \cite{Lan59}  the behavior in the vicinity of $x_{p'}^0$ should be
either logarithmic, proportional to $ |x_{p'}-x_{p'}^0|^\beta$, or
proportional to $|x_{p'}-x_{p'}^0|^\beta \log(x_{p'}-x_{p'}^0)$, where
$\beta$ is a noninteger number.

When $\mu\ll v$ we find from Eq.~(\ref{eq19d}):
\begin{equation}\label{20a} 
x_{p'}=\frac{1}{2}\pm \frac{v}{2v_{0}}
(1+\frac{\mu^4}{8v^4}).
\end{equation}
Comparing with Eq.~(\ref{eq3c}), we see that
at small $\mu$ or at large $v$ the singularities come 
closer to the physical region of $x_{p'}$.  We will see below that this 
will be a property of all the singularities depending on $\mu$ and 
$v$. This explains the numerical results, showing that with an increase 
of $v$ the graphs of the amplitudes versus $\alpha$ become more 
sharply peaked.
 
Now consider the case when one of the $\alpha$'s is zero. Let $\alpha_1=0$. 
Then Eq.~(\ref{eq3}) is reduced to:
\begin{eqnarray}
\varphi=\alpha_2(m^2-(p+q-k)^2) 
+\alpha_3(\mu^2-(p-k)^2)
\nonumber\\
\label{eq3e}
+\alpha_4(k^2-m^2)
+\gamma_1 \omega\cd (k-p').
\end{eqnarray}
Similarly to the previous case we get an equation, which can be obtained
from Eq.~(\ref{eq19b}) by deleting the first row and column:
\begin{equation}\label{eq13}
\left|
\begin{array}{lllll}
 2m^2    & \mu^2         &(s-2 m^2) &(1-x_{p'})\\ 
 \mu^2     & 2\mu^2        &-\mu^2      &(x_p-x_{p'}) \\
 (s-2m^2)& -\mu^2        & 2m^2     &  x_{p'}  \\ 
 (1-x_{p'}) & (x_p-x_{p'})   & x_{p'}      & 0  
\end{array} 
\right| =0.
\end{equation}
Under the given kinematical conditions
its solution with respect to $x_{p'}$ is
\begin{equation}\label{eq14}
x_{p'}^0=\frac{1}{2}\pm 
\frac{\mu}{4v}\sqrt{\frac{\mu^2+4v^2}{m^2+v^2}}.
\end{equation}
In the limit $\mu\rightarrow 0$ or $v\rightarrow \infty$ these singularities  
are again approaching the physical region.

Let $\alpha_2=0$. The equation is obtained from Eq.~(\ref{eq19b}) by 
deleting the second row and column:
\begin{equation}\label{eq15}
\left|
\begin{array}{lllll}
2\mu^2       &(2\mu^2-t)   & -\mu^2   & 0 \\ 
(2\mu^2-t)   & 2\mu^2       &-\mu^2      &(x_p-x_{p'}) \\
-\mu^2       & -\mu^2        & 2m^2     &  x_{p'}  \\ 
0            & (x_p-x_{p'})   & x_{p'}      & 0  
\end{array} 
\right| =0.
\end{equation}
Its solution reads:
\begin{equation}\label{eq16}       
x_{p'}^0=\frac{x_p\mu(4m^2\mu-\mu^3- \mu t
\pm 2\sqrt{t}\sqrt{tm^2+\mu^4-4m^2\mu^2})}
{4m^2\mu^2-(t-\mu^2)^2}.
\end{equation}
In the limit $v\rightarrow \infty$ it is simplified:
\begin{equation}\label{eq16a}       
x_{p'}^0=-\frac{\mu^2}{4v^2}\pm \frac{\mu m}{2v^2}.
\end{equation}

Let $\alpha_3=0$.
The equation is obtained from Eq.~(\ref{eq19b}) by deleting the third row
and column:
\begin{equation}\label{eq17}
\left|
\begin{array}{lllll}
2\mu^2       & \mu^2   & -\mu^2   & 0 \\ 
\mu^2        & 2m^2    &(s-2 m^2) &(1-x_{p'})\\ 
-\mu^2       & (s-2m^2)& 2m^2     &  x_{p'}  \\ 
0            & (1-x_{p'}) & x_{p'}      & 0  
\end{array} 
\right| =0.
\end{equation}
The determinant in Eq.~\r{eq17} can be evaluated:
\begin{equation}\label{eq17a}
4sx_{p'}^2-4sx_{p'}+4m^2-\mu^2=0.
\end{equation}
The solutions of Eq.~(\ref{eq17a}) are:
\begin{equation}\label{eq18}
x_{p'}^0=\frac{1}{2} \pm \frac{\sqrt{v^{2}+\mu^2/4}}{2v_{0}}.
\end{equation}
For $\mu\rightarrow 0$ they also approach the boundary of the physical 
region of $x_{p'}$.

Now consider the cases when a few coefficients are zero. 
Let $\alpha_1=\alpha_3=0$. 
The equation can be obtained from Eq.~(\ref{eq19b}) by deleting the 
first and third rows and columns: 
\begin{equation}\label{eq6a}
\left|
\begin{array}{lllll}
 2m^2           &(s-2 m^2) &(1-x_{p'})\\ 
 (s-2m^2)       & 2m^2     &  x_{p'}  \\ 
 (1-x_{p'})       & x_{p'}      & 0  
\end{array} 
\right| =0.
\end{equation}
This equation reduces to:
\begin{equation}\label{eq7}
x_{p'}^2-x_{p'} s+m^2=0.
\end{equation}
Its solutions are:
\begin{equation}\label{eq8}
x_{p'}^0=\frac{1}{2} + \frac{v}{2v_{0}}=x_{{\rm max}}, \; \;
x_{p'}^0=\frac{1}{2} - \frac{v}{2v_{0}}=x_{{\rm min}}.
\end{equation}
Since we consider the interval $0\leq \alpha\leq \pi$ corresponding to 
$x_{\rm min} \leq x_{p'} \leq1/2$, the singularity 
at $x_{p'}^0=x_{\rm max}$ is beyond the 
physical region, whereas the singularity at 
$x_{p'}^0=x_{\rm min}$ is just on the boundary of 
the physical region. The amplitude in this point gets an 
imaginary part:
\begin{eqnarray}
\label{eq1001}
{\rm Im}\;\Mone \; \neq 0 {\;\;\rm at \;\;} x_{p'}> x_{\rm min},\\ 
\label{eq1002}
{\rm Im}\;\Mone \;     =0 {\;\;\rm at \;\;} x_{p'}= x_{\rm min}.
\end{eqnarray}
Eq.~\r{eq1002} corresponds to $\alpha= \pi/2$. This explains, why all
the dashed curves of the imaginary parts in Fig.~\ref{figabove} go through zero
at the point $\alpha=\pi/2$.

Now put $\alpha_1=\alpha_2=0$. The corresponding 
equation is obtained from Eq.~(\ref{eq19b}) by deleting the first and
second rows and columns:
\begin{equation}\label{eq9}
\left|
\begin{array}{lllll}
 2\mu^2        &-\mu^2      &(x_p-x_{p'}) \\
 -\mu^2        & 2m^2     &  x_{p'}  \\ 
 (x_p-x_{p'})   & x_{p'}      & 0  
\end{array} 
\right| =0.
\end{equation}
Eq.~(\ref{eq9}) reads:
\begin{equation}\label{eq10}       
x_{p'}^2m^2-x_{p'} x_p(2m^2-\mu^2)+x_p^2m^2=0.
\end{equation}
Its solution is:
\begin{equation}\label{eq11}       
x_{p'}^0= \frac{x_{p}}{2m^2}\left(2m^2-\mu^2 
\pm i\mu\sqrt{4m^2-\mu^2}\right). 
\end{equation}
These two singularities are fixed points in the complex plane.
At $x_p=1/2$ and $\mu\ll m$ they are approaching the point 
$x_{p'}=1/2$ in the physical region, i.e., $\alpha=0$ and $\alpha=\pi$.

The case $\alpha_2=\alpha_3=0$ leads to the 
equation obtained from Eq.~(\ref{eq19b}) by deleting the second and
third rows and columns:
\begin{equation}\label{eq12}
\left|
\begin{array}{lllll}
2\mu^2      & -\mu^2   & 0 \\ 
-\mu^2      & 2m^2     &  x_{p'}  \\ 
0           & x_{p'}   & 0  
\end{array} 
\right| =0.
\end{equation}
It gives $x_{p'}^{0}=0$. This is a fixed singularity in the nonphysical 
region.

Above we have considered the region $ \omega\cd p \geq \omega\cd p'$.
In the region $ \omega\cd p \leq \omega\cd p' $ the integration domain
is restricted by the step function $\theta(\omega\cd (p-k))$ instead
of $\theta(\omega \cd (p'-k))$ in Eq.~(\ref{eq2}). 
The integrals defining these amplitudes define different analytic
functions depending on the region considered.
In the point $x_{p'}=1/2$, i.e. at $\alpha=0$ and $\alpha=\pi$, the
values of the functions coincide, but their analytic behavior is different.

This can indeed be seen in Fig.~\ref{figabove}. The slopes at $\alpha=0$ and 
$\alpha=\pi$ are different.

%%%%%%%%%%%%%%%%%%%%%%%%%%%%%%%%%%%%%%%%%%%%%%%%%%%%%%%%%%%%%

\subsection{Diamond}
\begin{figure}
\hspace{2.5cm} \epsfxsize=2.9cm \epsffile{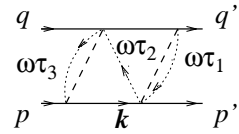}
\caption{\label{figdiamond}The diamond in explicitly covariant LFD.}
\end{figure}

The diamond corresponding to Eq.~\r{box2}
 is shown in Fig.~\ref{figdiamond}. 
The analytical expression is:
\begin{eqnarray}\label{eq23} 
\Mtwo &=&\int\theta(\omega\cd 
k)\delta(k^2-m^2) \nonumber \\ &&\times\theta(\omega\cd (k-p')) 
\delta((k-p'+\omega \tau_1 - \omega\tau_2)^2-\mu^2)
\nonumber \\           
&&\times\theta(\omega\cd (p+q-k)) \delta((p+q-k+\omega\tau_2)^2-m^2) 
\nonumber \\
&&\times\theta(\omega\cd (p-k))                                       
 \delta((p-k+\omega\tau_3)^2-\mu^2) \nonumber\\ 
 &&\times\frac{{\rm d}\tau_1{\rm d}\tau_2{\rm d}\tau_3} 
 {(\tau_1-i\epsilon)(\tau_2-i\epsilon)(\tau_3-i\epsilon)} 
 \frac{{\rm d}^4k}{(2\pi)^3}.
 \nonumber 
\\                                                                   
\end{eqnarray}                                                   
Performing the integrations in  Eq.~(\ref{eq23}) over $\tau_i$, we get:
\begin{eqnarray}\label{eq24} 
&&\Mtwo\nonumber = \\&&\int\frac{\theta(\omega\cd (k-p'))}{\mu^2-(k-p')^2
+\frac{\displaystyle{\omega\cd (k-p')}}{\displaystyle{\omega\cd 
(p+q-k)}}[m^2-(p+q-k)^2]}
\nonumber\\
&&\times\frac{\theta(\omega\cd (p+q-k))}{m^2-(p+q-k)^2} 
%\nonumber\\
\frac{\theta(\omega\cd (p-k))}{\mu^2-(p-k)^2}\ 
\theta(\omega\cd k)\frac{{\rm d}^3k}{2\varepsilon_k(2\pi)^3}.  
\end{eqnarray}

We still suppose that $\omega\cd p > \omega \cd p'$. However, now,
in contrast to the trapezium, $\omega \cd p' \leq \omega \cd k \leq 
\omega \cd p,$
and both restrictions have to be taken into account.

In order to find the singularities, one should consider the extremum of 
the function:
\twocolumn[\hsize\textwidth\columnwidth\hsize\csname
@twocolumnfalse\endcsname
\begin{eqnarray}\label{eq25} 
\varphi&=&\alpha_1\left\{\mu^2-(k-p')^2
+\frac{\displaystyle{\omega\cd (k-p')}}{\displaystyle{\omega\cd 
(p+q-k)}}(m^2-(p+q-k)^2)\right\}
\nonumber\\
&&+\alpha_2\left\{m^2-(p+q-k)^2\right\} 
+\alpha_3\left\{\mu^2-(p-k)^2\right\}
+\alpha_4(k^2-m^2) 
\nonumber\\
&&+\gamma_1\omega\cd (k-p')
+\gamma_2\omega\cd(k-p).
\end{eqnarray}  
At $\omega\cd p'=\omega\cd p$, i.e. at $\alpha=0$ and $\alpha=\pi$ the
integration domain vanishes and the diamond becomes zero, as shown in
Fig.~\ref{figabove}.  It remains zero in the interval $\pi \leq
\alpha\leq 2\pi$.

%%%%%%%%%%%%%%%%%%%%%%%%%%%%%%%%%%%%%%%%%%%%%%%%%%%%%%%%%%%%%%%%%%%

\subsection{Stretched box}
\begin{figure}
\hspace{6.5cm} \epsfxsize=2.9cm \epsffile{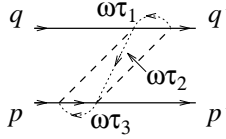}
\caption{\label{figsbox}The stretched box in explicitly covariant LFD.}
\end{figure}

The stretched box, corresponding to Eq.~\r{box0} is shown in 
Fig.~\ref{figsbox}. 
The analytical expression is:
\begin{eqnarray}\label{eq26} 
\Mthree&=&\int\theta(\omega\cd 
k)\delta(k^2-m^2) \nonumber \\ &&\times \theta(\omega\cd (k-p')) 
\delta((k-p'+\omega\tau_2-\omega\tau_3))^2 -\mu^2) \nonumber \\ 
&&\times\theta(\omega\cd (p+q-k)) \delta((p+q-k +\omega\tau_1 + 
\omega\tau_3 -\omega \tau_2)^2 -m^2) \nonumber \\ &&
 \times\theta(\omega\cd (p-k)) \delta((p-k+\omega \tau_3)^2 -\mu^2) 
\nonumber \\ &&\times\frac{{\rm d}\tau_1{\rm d}\tau_2{\rm d}\tau_3} 
 {(\tau_1-i\epsilon)(\tau_2-i\epsilon)(\tau_3-i\epsilon)} 
 \frac{{\rm d}^4k}{(2\pi)^3}\ .                                                   
\end{eqnarray}                                                                    
Performing the integrations in  Eq.~(\ref{eq26}) over $\tau_i$, we get:
\begin{eqnarray}\label{eq27} 
\Mthree&=&\int
\frac{\theta(\omega\cd (p+q-k))}{m^2-(p+q-k)^2
+\frac{\displaystyle{\omega\cd (p+q-k)}}{\displaystyle{\omega\cd 
(k-p')}}[\mu^2-(k-p')^2]}
\nonumber\\
&&\times\frac{\theta(\omega\cd (k-p'))}{\mu^2-(k-p')^2+
\frac{\displaystyle{\omega\cd (k-p')}}{\displaystyle{\omega\cd 
(p-k)}}[\mu^2-(p-k)^2]} 
%\nonumber\\ &&\times
\frac{\theta(\omega\cd (p-k))}{\mu^2-(p-k)^2}\ 
\theta(\omega\cd 
k)\frac{{\rm d}^3k}{2\varepsilon_k(2\pi)^3}.  
\end{eqnarray}

In order to find the singularities, one must consider the extremum of 
the function:
\begin{eqnarray}\label{eq28} 
\varphi&=&\alpha_1\left\{m^2-(p+q-k)^2
+\frac{\displaystyle{\omega\cd (p+q-k)}}{\displaystyle{\omega\cd 
(k-p')}}(\mu^2-(k-p')^2)\right\}
\nonumber\\
&&+\alpha_2\left\{\mu^2-(k-p')^2+
\frac{\displaystyle{\omega\cd (k-p')}}{\displaystyle{\omega\cd 
(p-k)}}(\mu^2-(p-k)^2)\right\}
\nonumber\\
&&+\alpha_3\{ \mu^2-(p-k)^2\} +\alpha_4(k^2-m^2) 
%\nonumber\\ &&
+\gamma_1\omega\cd (k-p')+\gamma_2\omega\cd(k-p).
\end{eqnarray}          

Calculating the derivative of Eqs.~(\ref{eq3e}), (\ref{eq25}) and
(\ref{eq28}), for example, with respect to $\alpha_{1-4}$, at
$\gamma_{1}=\gamma_{2}=0$, one finds identical equations determining
the singularities. These do not depend on $x_{p},x_{p'}$ and coincide
with the ones of the Feynman graph.  Similarly one can see that any
singularity depending on $x_{p}$ and $x_{p'}$ cannot appear in a
separate diagram only. It appears at least in two amplitudes. These
singularities cancel each other in the sum of the amplitudes.
%\vspace{1.5cm}
%\clearpage
]

%%%%%%%%%%%%%%%%%%%%%%%%%%%%%%%%%%%%%%%%%%%%%%%%%%%%%%%%%%%%%%%

\section{Analysis of the off energy-shell results}
\label{secoffshell}
\begin{figure}
\hspace{-.2cm}
\epsfxsize=8.5cm \epsffile{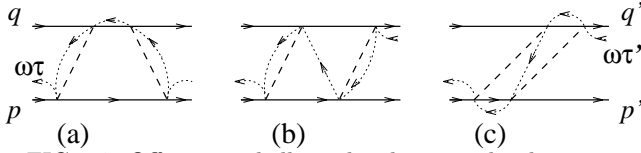}
\caption{\label{figkoff}Off energy-shell amplitudes in explicitly covariant
LFD: (a) The trapezium. (b) The diamond. (c) The stretched box. The external
momenta are the same for all diagrams.}
\end{figure}

The off energy-shell amplitude is shown graphically in Fig.~\ref{figkoff}.  
It contains incoming
and outgoing spurion lines with the momenta $\omega\tau$ and $\omega\tau'$,
respectively.  The conservation law has the form:
\begin{equation}\label{eq29}
p+q-\omega\tau = p'+q'-\omega\tau' = P.
\end{equation}
From Eq.~\r{eq29} one can infer that if $\vec{P}=0$, then $\vec{p'} + \vec{q'} \neq 0$, as
it was indicated for the $z$-components in Sec.~\ref{secundert}.
To parametrize the off energy-shell amplitude, we introduce different
initial and final Mandelstam variables $s$:
\begin{equation} s=(p+q)^2,\quad s'= (p'+q')^2, \end{equation}
and the total mass squared:
\begin{equation}  {\cal M}^2= (p+q-\omega\tau)^2= (p'+q'-\omega\tau')^2. \end{equation}
So, in general, the off energy-shell amplitude is parametrized as:
\begin{equation}\label{0}
M=M(s,s',{\cal M}^2,t,x_p,x_{p'}).
\end{equation}
The on energy-shell amplitude Eq.~(\ref{eq1a}) is obtained from Eq.~(\ref{0})
by the substitution $s=s'={\cal M}^2$. One can also consider the half off energy-shell
amplitude with one incoming or outgoing spurion line. It is obtained
from Eq.~(\ref{0}) by the substitutions $s={\cal M}^2\neq s'$ or $s'={\cal M}^2\neq s$.

In the case of the trapezium, Fig.~\ref{figkoff}a, the external spurion
lines enter and exit from the diagram at the same points as the momenta
$p$ and $p'$. So, they can be incorporated by the replacement:
\begin{equation}
p \rightarrow p-\omega\tau,\quad p'\rightarrow p'-\omega\tau'.
\end{equation}
This
corresponds to new masses of the initial and final particles for the
bottom line:

\begin{eqnarray}
m_i^2=(p-\omega\tau)^2=m^2-x_p(s-{\cal M}^2), \nonumber\\
m_f^2=(p'-\omega\tau')^2=m^2-x_{p'}(s'-{\cal M}^2). 
\end{eqnarray}
With these new masses, one can repeat the calculations of Sec.~\ref{trap} 
and find the singularities of the off energy-shell amplitude
for the trapezium. The masses of the intermediate particles are not
changed.

For other diagrams, both for the diamond and the stretched box, in contrast to 
the 
trapezium,  the spurion line enters in the point where the momentum $q'$
go out from the graph. This means that the calculation has to be done with
the following external mass of this particle:
\begin{equation}m'^2=m^2\rightarrow (q'-\omega\tau)^2=m^2-(1-x_{p'})(s'-{\cal M}^2),\end{equation}
whereas the mass of the particle with momentum $p'$ is $m$.

Like in the case when all masses are equal, the sum of all time-ordered
graphs with masses different from the internal ones, but 
the same in all the time-ordered graphs, would not depend on $\omega$.  
However,
now we take the sum of the graphs with different external masses in
different particular graphs. This sum cannot be obtained by the time-ordering 
of a given Feynman graph. In this case the $\omega$-dependence is
not eliminated in the sum of all the graphs, and the exact off
energy-shell amplitude in a given order still depends on $\omega$.
An example of this dependence is shown in Fig.~\ref{figoff}.

The off-shell amplitude is not a directly observable quantity. It may
enter as part of  a bigger diagram. Therefore, the off shell 
amplitude may depend on $\omega$. This $\omega$-dependence is not forbidden
by covariance and, hence, does not violate it. On energy-shell this
dependence disappears.

\section{\label{secconc} Conclusions}
 
If sufficient caution is exercised, invariance of $S$-matrix elements can be
maintained in Hamiltonian formulations of field theory. A necessary condition
to be fulfilled is that all Fock sectors included in the Feynman diagrams that
contribute to a perturbative approximation of the $S$-matrix are retained.
For the specific case of scalar field theory at fourth order in the coupling
constant we have determined the magnitude of the breaking of covariance if only
the diagrams generated by the ladder approximation to Hamiltonian dynamics are
included. The remaining terms, the stretched boxes, were found to contribute a 
small fraction, less than 2\% for small to intermediate c.m.s. momenta, of the
total amplitude. This fraction is, however, increasing with energy.

It was found, in a calculation closely approximating the first one, that the
violation of invariance is much larger in instant-form dynamics than in 
light-front dynamics, confirming quantitatively what has been claimed in the
literature.

In both  cases we determined quantitatively the dependence of the six LF
time-ordered diagrams on the orientation of the light-front. We verified that,
although the individual diagrams depend strongly on the orientation, their sum
does not, as it should not. This dependence of individual diagrams may be
interpreted as a breaking of rotational invariance.

Having established numerically that invariance of the $S$-matrix
elements is obtained only if all Fock sectors relevant to a certain
order in perturbation theory are included, we extended our
investigation to amplitudes that are off-energy-shell. Such amplitudes
are not $S$-matrix elements, calculated between asymptotic states, from
$-\infty$ to $+\infty$ in time. They are elements of an $S$-matrix
calculated for finite light-front time, i.e., defined on a light-front
in the interaction region, not moved to $\pm\infty$ \cite{CDKM98}.
Therefore they depend on the orientation of this light-front. They
either occur as parts of larger diagrams that are invariant, or in the
calculation of LF wave functions. Not being invariant, the sum of the
six LF time-ordered diagrams corresponding to the box is expected to
depend on the orientation of the light-front. We found the variation of
the sum of these six diagrams to grow more strongly with increasing
relative momentum than the fraction carried by the stretched boxes.

All these results point to the conclusion that for low and intermediate
momenta, e.g. those relevant for the bulk of the deuteron wave
function, the higher Fock components are very small and are expected to
play a minor role in LF dynamics. We conjecture that this conclusion
remains essentially valid for higher orders in perturbation theory.

Two remarks are in order here. First, if bound or scattering states at high
values of the relative momentum are to be calculated, the higher Fock states 
will become much more important. Secondly, in the present work we neglected
spin. It remains to be seen to what extent the special effects of spin, notably
instantaneous propagators, will influence our conclusions.

A final point concerns the dependence of the individual diagrams on the 
orientation of the light-front. By an analysis very close to the Landau method
for Feynman diagrams, we were able to explain all the peculiarities of the
angular dependence in terms of the occurrence and position of singularities of 
the time-ordered diagrams as a function of the angles and their locations. In
particular the symmetries of the angular dependence and the cusps showing up at
specific orientations could be explained fully.

\section*{Acknowledgements}
 
The authors thank A.J.~Poldervaart for writing the first version of the
{\sc fortran} code used. This work was supported by
the Stichting voor Fundamenteel Onderzoek der Materie (FOM), which is
financially supported by the Nederlandse Organisatie voor
Wetenschappelijk onderzoek (NWO).

\end{document}